\documentclass[prl,twoside,preprintnumbers,superscriptaddress,twocolumn,nofootinbib,nobibnotes]{revtex4}
\pdfoutput=1 
\usepackage{amsmath,graphicx,slashed,multirow,color,ulem}
\usepackage{graphicx,epsfig,amssymb,bm,slashed,wasysym,multirow}
\usepackage[utf8]{inputenc}
\def \beq{\begin{equation}}
\def \eeq{\end{equation}}
\def \beqa{\begin{eqnarray}}
\def \eeqa{\end{eqnarray}}

\def\msbar{$\overline{\hbox{MS}}$}

\def\bea{\begin{eqnarray}}
\def\eea{\end{eqnarray}}

\preprint{IPPP/16/51}
\begin{document}
\title{QCD radiative corrections for $h\to b\bar b$ in the Standard Model Dimension-6 EFT}

\author{Rhorry Gauld}
\email{rhorry.gauld@durham.ac.uk}
\affiliation{Institute for Particle Physics Phenomenology, University of Durham, DH1 3LE Durham, United Kingdom}

\author{Benjamin~D.~Pecjak}
\email{ben.pecjak@durham.ac.uk}
\affiliation{Institute for Particle Physics Phenomenology, University of Durham, DH1 3LE Durham, United Kingdom}

\author{Darren~J.~Scott}
\email{d.j.scott@durham.ac.uk}
\affiliation{Institute for Particle Physics Phenomenology, University of Durham, DH1 3LE Durham, United Kingdom}

\date{July 2016}

\begin{abstract} 
  We calculate the $\mathcal{O}(\alpha_s)$ QCD corrections to the
  inclusive $h\to b\bar b$ decay rate in the dimension-6 Standard
  Model Effective Field Theory (SMEFT).
  The QCD corrections multiplying the dimension-6 Wilson coefficients which alter
  the $hb\bar b$-vertex at tree-level are proportional to the Standard Model (SM) ones,
  so next-to-leading order results can be obtained through a simple rescaling 
  of the tree-level decay rate.
  On the other hand, contributions from the operators $Q_{bG}$ and
  $Q_{HG}$, which alter the $gb\bar b$-vertex and introduce a $hgg$-vertex
  respectively, enter at $\mathcal{O}(\alpha_s)$ and induce sizeable
  corrections which are unrelated to the SM ones and
  cannot be anticipated through a renormalisation-group analysis.
  We present compact analytic results for these contributions, which
  we recommend to be included in future phenomenological studies.

\end{abstract}

\maketitle

\section{Introduction}

The precise determination of the properties of the Higgs-like particle
discovered during Run-I of the
LHC~\cite{Aad:2012tfa,Chatrchyan:2012ufa} is a main focus of the
Run-II physics program.
Within the current experimental precision, the coupling strengths of
this particle appear to be consistent with the Standard Model (SM)
Higgs boson~\cite{Bolognesi:2012mm,Aad:2013xqa,Aad:2015gba,Khachatryan:2014jba,ATLAS-CONF-2015-044}.
With both theoretical and experimental developments, a significant improvement in the precision of the determination
of the Higgs couplings is expected at the LHC during the High Luminosity (HL) phase~\cite{Dawson:2013bba}.

The determination of the Higgs couplings which enter both
production and decay processes is achieved by performing a
(global) fit to $\sigma \cdot BR(H\to X)$ data of the observable final states.
This is necessary since a direct measurement of the SM Higgs boson width 
is not feasible at the LHC (or at proposed future colliders) since the expected value of 
$\approx4$~MeV~\cite{Heinemeyer:2013tqa} is orders of magnitude
smaller than the mass resolution of the detectors.
The importance of a precision measurement of the $h\to b\bar b$ partial
width cannot be overstated, since this constitutes the
dominant decay channel for the SM Higgs boson
($\approx60\%$~\cite{Heinemeyer:2013tqa}). Consequently, a
modification of the Higgs coupling to $b$-quarks can have a sizeable
impact on the extraction of all other Higgs couplings as the Higgs
production rate and all branching fractions are modified by a shift
in the total decay width $\Gamma_h$.

Unfortunately, precision extractions of the $h\to b\bar b$ partial and total width
at the LHC are challenging.
A measurement of the $h\to b\bar b$ partial width is ultimately
limited at the LHC, since knowledge of both the shape and
normalisation of the contributing QCD backgrounds must be known to
high accuracy~\cite{Chatrchyan:2013zna,Aad:2014xzb}. Current
projections for a measurement of the $h\to b\bar b$ signal strength
indicate that precision $\simeq (5-7)\%$ may be achievable in the HL
phase of the LHC~\cite{Dawson:2013bba}.
An extraction of the total width from
experimental data (under minimal assumptions) through interference
effects~\cite{Dixon:2013haa,Caola:2013yja,Campbell:2013una,Campbell:2013wga,Campbell:2014gua}
is possible, and additional information can also be gained by including LEP
data~\cite{Englert:2015bwa}. However, the current constraint from this
technique is approximately $\Gamma_h < 5\,\Gamma_h^{SM}$~\cite{Khachatryan:2014iha,Aad:2015xua}. 
A precise model independent measurement of both these quantities
requires input from a dedicated `Higgs machine'.

There are several design proposals for a collider with such
capabilities, such as a linear/circular $e^+e^-$
machine~\cite{Aicheler:2012bya,Behnke:2013xla},
muon-collider~\cite{Alexahin:2013ojp,Delahaye:2013jla}, or
$\gamma\gamma$ machine~\cite{Asner:2001ia,Bogacz:2012fs}. Although the
proposed physics programs are different in each case, the common goal
is to achieve $\mathcal{O}(\%)$ level precision on the measurement of
Higgs couplings.
For example, through a combination of a Higgs recoil measurement and
an exclusive measurement of $h\to VV$, better than 10\% precision on
the total width can be expected~\cite{Baer:2013cma}. In the case of
the $h\to b\bar b$ branching fraction, better than $\mathcal{O}(\%)$
precision is expected in some cases --- see Table 2.3
of~\cite{Baer:2013cma}.  In such a scenario, precision calculations
are also required to interpret the data. In this work, we focus on the
QCD corrections to the Higgs decay to $b$-quarks in the framework of
the Standard Model Effective Field Theory (SMEFT).

In the SMEFT framework, the effects of physics beyond the SM are
parameterised by a set of non-vanishing Wilson coefficients of
higher-dimensional operators.  Practically, the usual dimension-4 SM
Lagrangian is extended to include operators of mass-dimension $n>4$,
which are constructed from gauge-invariant combinations of SM fields
and multiplied by Wilson coefficients of mass dimension ($4-n$).
These operators effectively describe the interactions of new physics
particles with those present in the SM, an approach which may be
justified if the energy scale of these new particles $\Lambda_{\rm
  NP}$ greatly exceeds the electroweak scale.  One of the main
benefits of this approach is that new physics effects can be
characterised without specifying a particular UV complete model
of physics Beyond-the-SM (BSM). The interpretation of data in terms of
non-vanishing Wilson coefficients is therefore performed in a
model-independent fashion.  In the absence of any direct evidence for
new particle states during Run-I of the LHC, we believe this approach
to be both justified and well motivated.

During Run-I, the interpretation of Higgs measurements by ATLAS and
CMS was generally performed in the `interim' $\kappa$ and signal
strength formalisms --- see for example the combined Run-I analysis of
CMS and ATLAS data~\cite{ATLAS-CONF-2015-044}. In Run-II, it is a
recommendation of the LHC Higgs Cross Section Working Group to move
towards a more general
EFT framework (see for instance Section 10.4 of~\cite{Heinemeyer:2013tqa}).  In doing
so, it is important to note that the predictions for observables
obtained in SMEFT are not unlike those obtained in the SM (or
UV-completions of the SM for that matter) where a perturbative
expansion has been applied --- higher-order corrections reduce the
theoretical uncertainty of the predictions of observables. Moreover,
the next-to-leading order (NLO) corrections to a given observable
typically depend on Wilson coefficients which are not present in the
tree-level result.  It is therefore important to extend SMEFT analyses
to NLO (and beyond), to allow for a more precise determination of Wilson
coefficients (and allowed ranges) through a comparison with
experimental data, and much recent work has been dedicated to this
task for a wide range of 
processes~\cite{Elias-Miro:2013gya,Elias-Miro:2013mua,
  Elias-Miro:2013eta,Jenkins:2013zja,Zhang:2013xya,Jenkins:2013wua,Alonso:2013hga,
  Passarino:2012cb,Chen:2013kfa,Grojean:2013kd,Zhang:2014rja,Englert:2014cva,Zhang:2014lya,Pruna:2014asa,Henning:2014wua,Degrande:2014tta,Franzosi:2015osa,Ghezzi:2015vva,
  David:2015waa,
  Grober:2015cwa,Hartmann:2015oia,Hartmann:2015aia,Gauld:2015lmb,Zhang:2016omx,Bylund:2016phk,Gorbahn:2016uoy,Maltoni:2016yxb}.

At present, it is possible to deduce logarithmically enhanced NLO
corrections appearing in fixed-order perturbation theory to arbitrary
observables using renormalisation-group (RG)
equations for the Wilson coefficients along with the full one-loop
anomalous dimension matrix calculated in
\cite{Jenkins:2013zja,Jenkins:2013wua,Alonso:2013hga}.  However, it is
common practice to use RG-improved perturbation theory to absorb such
logarithmic corrections into the running of the scale-dependent Wilson
coefficients between $\Lambda_{\rm NP}$ and the scale at which the
underlying decay or scattering process takes place (for $h\to
b\bar{b}$ decays this is $m_h$).  This removes large logarithms involving $\Lambda_{\rm NP}$
from effective theory matrix elements, and allows constraints on Wilson
coefficients obtained at experimentally accessible energy scales to be
interpreted at the scale $\Lambda_{\rm NP}$ where the effective
interactions are generated.  The remaining NLO corrections do not
contain large logarithms involving the scale $\Lambda_{\rm NP}$, and
cannot be deduced from an RG analysis.  However, these corrections can still be 
important numerically because for the interesting region of 
$\Lambda_{\rm NP}\sim 1$~TeV the RG-induced logarithms are not
dramatically enhanced.

In this work, we extend our previous calculation of such NLO SMEFT
corrections arising from four-fermion contributions and the
(presumably) numerically dominant electroweak
corrections~\cite{Gauld:2015lmb} by computing the
$\mathcal{O}(\alpha_s)$ correction to the $h\to b\bar b$ decay
rate. We proceed by introducing the relevant details of the SMEFT
framework for the $h\to b\bar b$ decay, and discuss the
renormalisation procedure we adopt for performing SMEFT NLO
calculations. We then provide the analytic results, and make
recommendations for their use in phenomenological studies.

\section{Calculational Set-up}\label{SetUp}
\subsection{Preliminaries}
In the SMEFT, the usual SM Lagrangian is appended by
higher-dimensional operators multiplied by Wilson coefficients. In the
current work we are interested in dimension-6 operators, and so use a
Lagrangian of the form
\begin{align}
\label{eq:Lagrangian}
{\cal L}={\cal L}_{\rm SM} + {\cal L}^{(6)} \, ; \qquad {\cal L}^{(6)} = 
\sum_i C_i(\mu) Q_i(\mu) \,.
\end{align} 
The operators relevant for this work are listed in
Table~\ref{op59}. Note that in our convention the  Wilson
coefficients of dimension-6 operators have mass dimension minus two, so that the $C_i$ are
suppressed by $\Lambda_{\rm NP}^2$, and that we have rescaled the
operator $Q_{dG}$ by a factor of the strong coupling constant $g_s$ with respect to the usual
definition.  The relevant interaction Lagrangian for Higgs couplings
to down-type quarks is
\begin{table}[t]
\begin{center}
\small
\renewcommand{\arraystretch}{1.5}
\begin{tabular}[t]{c|c}
\hline
$Q_{H\Box}$ & $(H^\dag H)\Box(H^\dag H)$ \\
$Q_{H D}$   & $\ \left(H^\dag D_\mu H\right)^* \left(H^\dag D_\mu H\right)$ \\
\hline
$Q_{dH}$           & $(H^\dag H)(\bar q_p d_r H)$\\
\hline
$Q_{H G}$     & $H^\dag H\, G^A_{\mu\nu} G^{A\mu\nu}$ \\
$Q_{H\widetilde G}$         & $H^\dag H\, \widetilde G^A_{\mu\nu} G^{A\mu\nu}$ \\
\hline
$Q_{dG}$        & $g_s (\bar q_p \sigma^{\mu\nu} T^A d_r) H\, G_{\mu\nu}^A$ \\
\end{tabular}
\end{center}
\caption{\label{op59}
A sub-set of the 59 independent dimension-6 operators built from Standard Model fields which conserve baryon number relevant for the current calculation, as given in Ref.~\cite{Grzadkowski:2010es}. The subscripts $p,r$ are flavour indices,
and $q_p$ and $d_r$ are left- and right-handed fields, respectively.
}
\end{table}
\begin{align} \label{hff}
\mathcal{L}_{\rm Higgs} =& -\left[  H^{\dagger } \overline d_r  [Y_d]_{rs} \,   q_{s} +  h.c. \right] 
\nonumber \\[1mm] &
+\left[  C_{\substack{d H \\ sr}}^* (H^\dagger H) H^{\dagger } \overline d_r \, q_{s} +h.c.\right] \, .
\end{align}
where $[Y_d]$ and $C_{\substack{d H}}$ are complex matrices in flavour space. 
The Higgs potential is also altered compared to the SM, and requiring that the 
kinetic terms are canonically normalised leads one to write the Higgs doublet 
in unitary gauge in the broken phase of the theory as 
\begin{align} \label{eq:Higgs}
 H(x) = \frac{1}{\sqrt{2}}\left( \begin{array}{c}
0 \\
 \left[1+C_{H,{\rm kin}}\right]h(x) 
 + v_T\end{array}  \right),
 \end{align}
where
\begin{align} \label{eq:Class3}
C_{H,{\rm kin}} \equiv \left(C_{H\Box}-\frac{1}{4}C_{HD}\right)v_T^2  \,  ,
\end{align}
and the vacuum expectation value $v_T\approx (\sqrt{2} G_F)^{-\frac{1}{2}}$, with $G_F$ 
the Fermi constant.
It should be noted that although $C_{H, \rm kin}$ is dimensionless, due to the presence of $v_T^2$ in its definition, it is still understood to be implicitly suppressed by $\Lambda_{\rm NP}^{2}$.

Throughout this work, we are concerned with the flavour conserving process $h\to b\bar b$.
However, the matrices $[Y_d]$ and $C_{\substack{d H}}$
appearing in Eq.~(\ref{hff}) are in general not simultaneously diagonalisable, leading to
flavour violating effects which are not present in the SM. The running of $C_{\substack{d H}}$ 
also introduces further flavour violating effects which should be considered~\cite{Alonso:2013hga,Wells:2015cre}.
We follow the procedure taken in previous work~\cite{Gauld:2015lmb},
imposing a minimal flavour violation (MFV) scenario~\cite{Chivukula:1987py,D'Ambrosio:2002ex} 
and setting $V_{tb} = 1$ throughout.
The $b$-quark Yukawa coupling, defined as the coefficient of the 
$h b\bar b$-vertex in the mass basis of the broken phase of
the theory, is therefore related to the physical mass according to
\begin{align}\label{eq:YMrel}
y_b = \sqrt{2}\frac{m_{b}}{v_T} + \frac{v_T^2}{2} C_{bH}^* \,.
\end{align} 
With these preliminaries in place, it is straightforward to compute the tree-level
amplitude for the process $h\to b\bar b$ as
\begin{align}
i{\cal M}^{\rm tree} = 
-i \bar u(p_b) 
\left({\cal M}_{b,L}^{{\rm tree}} P_{L} + {\cal M}_{b,L}^{{\rm tree}*} P_R\right)v(p_{\bar b}) \, ,
\end{align}
where
\begin{align}
\label{eq:Mtree}
{\cal M}_{b,L}^{{\rm tree}} &=  
\frac{m_b}{v_T}\left[ 1+C_{H,{\rm kin}} \right]-  \frac{ v_T^2}{\sqrt{2}}C^{*}_{bH} \, .
\end{align}

\subsection{Renormalisation  procedure}
The calculation of the QCD corrections to $h\to b\bar b$ in the SMEFT 
proceeds much the same way as in the SM.  In addition to calculating the 
real and virtual contributions to the NLO matrix elements, which we discuss 
below, we must also construct a set of counterterms which render the 
virtual corrections UV finite.    There is some subtlety in the
construction of such UV counterterms, and we therefore provide details on
the renormalisation procedure here\footnote{The reader is directed to a 
previous publication~\cite{Gauld:2015lmb} for a more extensive discussion.}.
In essence, wavefunction and parameter/mass renormalisation is performed 
in the on-shell scheme, and Wilson coefficients are renormalised in the \msbar~scheme.

We first discuss the calculation of the renormalisation constants for the
external fields and parameters/masses.  
This proceeds  as in the SM, and a detailed discussion of this procedure for the SM can
be found in~\cite{Denner:1991kt}.  While the full procedure including also electroweak 
corrections is rather involved, for the QCD corrections to $h\to b\bar{b}$ we need only
renormalise the $b$-quark field and mass.  We relate renormalised and bare quantities
according to 
\begin{align}
\nonumber
b^{(0)}_{L,R} & = \sqrt{Z^{L,R}_b} b_{L,R} = \left(1+\frac12 \delta Z^{L,R}_b\right)b_{L,R} \, , \\
m_b^{(0)} & = m_b + \delta m_b\, ,
\end{align}
where the superscript $(0)$ labels the bare field or mass.  Explicit expressions for the 
one-loop renormalisation constants are obtained from the $b$-quark two-point function.
When computing the renormalisation constants, the divergences are
regulated by performing the loop integrals in $d = 4 - 2\epsilon$ dimensions. 
The relevant one-loop renormalisation constants are found to be
\begin{align}
\label{eq:4fctsB}
\nonumber
\frac{\delta m_b^{(6)}}{m_b} &=  -\frac{\alpha_s C_F}{\pi} \frac{m_b v_T}{2\sqrt{2}}   \left( 3\frac{C^b_{\epsilon}}{\hat{\epsilon}} + 1 \right) 
\left(C_{bG}+C_{bG}^* \right)\, ,
\nonumber \\[1mm]
\delta Z_b^{(6),L} &= \frac{\alpha_s C_F}{\pi} \frac{m_b^2 v_T}{4\sqrt{2}}\left( 3\frac{C^b_{\epsilon}}{\hat{\epsilon}} + 1 \right)
 \left(C_{bG}-3 C_{bG}^* \right) \, ,
\nonumber \\[1mm]
\delta Z_b^{(6),R} &= -\frac{\alpha_s C_F}{\pi} \frac{m_b^2 v_T}{4\sqrt{2}}\left(3 \frac{C^b_{\epsilon}}{\hat{\epsilon}} + 1 \right) 
\left(C_{bG}+C_{bG}^* \right) \, ,
\nonumber \\[1mm]
\frac{\delta m_b^{(4)}}{m_b} &=  -\frac{\alpha_s C_F}{\pi} \left( \frac{3}{4}\frac{C^b_{\epsilon}}{\hat{\epsilon}} + 1 \right) \, ,
\nonumber \\[1mm]
\delta Z_b^{(4),L} &= \delta Z_b^{(4),R} = \delta Z_b^{(4),L*} = \delta Z_b^{(4),R*} \, ,
\nonumber \\[1mm]
\delta Z_b^{(4)} &= 2 \delta Z_b^{(4),L} = -\frac{\alpha_s C_F}{\pi} \left( \frac{3}{2}\frac{C^b_{\epsilon}}{\hat{\epsilon}}  +2 \right) \, ,
\end{align}
where $N_c=3$ is the number of colours in QCD, $C_F=(N_c^2-1)/(2N_c)$, and
\begin{align}
C^b_{\epsilon} = 1+\epsilon \ln\left[\frac{\mu^2}{m_b^2}\right] \,,\qquad \frac{1}{\hat{\epsilon}} \equiv \frac{1}{\epsilon} - \gamma_E + \ln(4\pi) \,.
\end{align}
The SM and dimension-6 contributions are distinguished through the superscript $(4)$ and $(6)$
respectively. It is worth noting that we follow the convention of~\cite{Kniehl:1996bd} by requiring 
$\delta Z_b^{R}$ to be real.

We must also include counterterms related to operator renormalisation.  These counterterms are 
generated from the operators whose Wilson coefficients appear in the tree-level expression in Eq.~(\ref{eq:Mtree}),
and are in fact simple to construct using that expression as a starting point. 
We do this by interpreting those as bare Wilson coefficients, and replace them in the \msbar~scheme by 
renormalised coefficients according to 
\begin{align}
C_i^{(0)} = C_i(\mu) + \delta C_i(\mu) = C_i(\mu)+\frac{1}{2\hat{\epsilon}}\frac{\dot{C}_i(\mu)}{(4\pi)^2} 
 \,,
\end{align}
where
\begin{align}
\label{eq:delCdef}
\frac{\dot{C}_i(\mu)}{(4\pi)^2} \equiv  \mu \frac{d}{d\mu}C_i(\mu)  \,.
\end{align}
\begin{figure*}[t]
\centering
\includegraphics[scale=1.0]{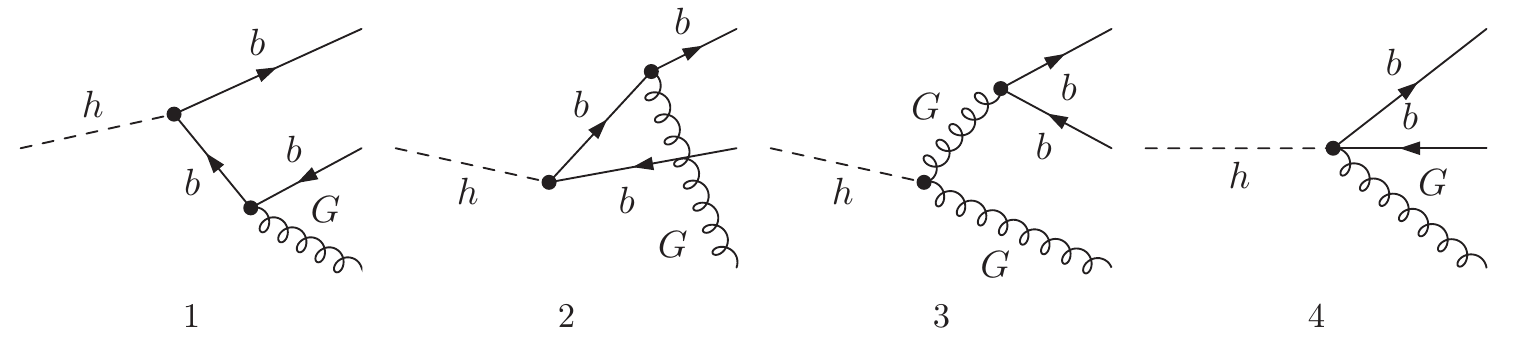} \\
\includegraphics[scale=1.0]{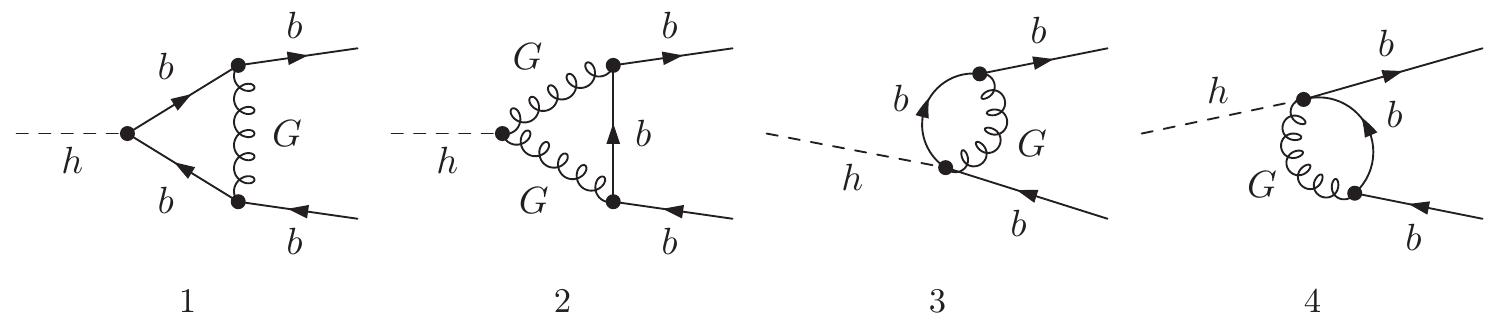}
\caption{Feynman diagrams contributing to both real (top) and virtual (bottom) $\mathcal{O}(\alpha_s)$ corrections to the $h\to b\bar{b}$ decay rate. The real corrections labelled $3$ and $4$ are generated by $Q_{HG}$ and $Q_{bG}$ operators respectively. Similarly, the virtual corrections labelled $2$ and $3,4$ are generated by $Q_{HG}$ and $Q_{bG}$ operators respectively.}
\label{fig:Hbb}
\end{figure*}
Explicit expressions for the $\delta C_i$ can be obtained
from the one-loop anomalous dimension calculation in the unbroken phase of the theory performed in~\cite{Jenkins:2013zja,Jenkins:2013wua,Alonso:2013hga}. Where necessary, these results are converted
into the broken phase using the tree-level SM relations, a procedure which is consistent to 
$\mathcal{O}(\Lambda_{\rm NP}^{-2})$. Extracting only the relevant results for the QCD corrections, we find
\begin{align}
\label{delCbh}
\nonumber
&\delta C_{bH} =\frac{\alpha_s C_F}{\pi}\frac{3}{v_T^2}\frac{1}{\hat{\epsilon}} \\
&\phantom{xxx} \times \bigg(2m_b^2C_{bG}  + v_T\left( \sqrt{2} m_b C_{HG} - \frac{v_T}{4} C_{bH} \right)  \bigg) \,.
\end{align}
The corresponding counterterm for $C_{H,{\rm kin}}$ at ${\cal O}(\alpha_s)$ 
is zero.  Moreover, we have omitted a term proportional to $C_{H\tilde{G}}$, which contributes to the 
counterterm amplitude but not to the decay rate at ${\cal O}(\alpha_s \Lambda_{\rm NP}^{-2})$ .

Having provided all the necessary renormalisation constants,
the counterterm amplitude can be constructed, and is generally written as
\begin{align} \label{CT}
{\cal M}^{\rm C.T.}(h\to b\bar b)= 
-\bar u(p_b) 
\left(\delta {\cal M}_L P_L + \delta {\cal M}_L^*  P_R\right) v(p_{\bar b}) \,.
\end{align}
The expression for the (real) SM counterterm is
\begin{align}
\label{eq:dim6CT}
\delta {\cal M}_L^{(4)} = &\frac{m_b}{v_T}\left(\frac{\delta m_b^{(4)}}{m_b} + \delta Z_b^{(4)}\right)  \,,
\end{align}
and the corresponding dimension-6 counterterm is
\begin{align}
\label{eq:dim6CT}
\delta {\cal M}_L^{(6)} =& 
\left( \frac{m_b}{v_T} C_{H,{\rm kin}} \right)  \left(\frac{\delta m_b^{(4)}}{m_b} 
+ \delta Z_b^{(4)}\right) \nonumber \\ 
&- \frac{v_T^2}{\sqrt{2}}   \left(  \delta C^*_{bH} +  C^*_{bH}\delta Z_b^{(4)}\right) 
 \nonumber \\
&+\frac{m_b}{v_T}\left(\frac{\delta m_b^{(6)}}{m_b} + \frac12 \delta Z_b^{(6),L}+ \frac{1}{2} \delta Z_b^{(6),R*}\right)  \,.
\end{align}

\section{Results for decay rate}
\label{Results}
The differential decay rate to NLO is obtained by evaluating the expression
\begin{align}
\label{eq:dGamma}
d\Gamma = \frac{d \phi_2}{2 m_h}\sum \big{|} {\cal M}_{h\to b\bar b}  \big{|}^2
	+\frac{d \phi_3}{2 m_h} \sum\big{|} {\cal M}_{h\to b\bar bg}  \big{|}^2 \,,
\end{align}
where $d\phi_i$ is the $i$-body differential phase-space factor.
The UV finite two-body contribution is defined by
\begin{align}
{\cal M}_{h\to b\bar b} = {\cal M}^{\rm one-loop} + {\cal M}^{\rm C.T.} +  {\cal M}^{\rm tree} \,,
\end{align}
and the three-body term ${\cal M}_{h\to b\bar bg}$ is the real emission amplitude.  

To compute real and virtual amplitudes (squared), the SM and the relevant 
dimension-6 Lagrangian have been implemented in  \texttt{FeynRules}~\cite{Alloul:2013bka}.
The contributing Feynman diagrams are subsequently generated and computed with 
\texttt{FeynArts}~\cite{Hahn:2000kx}  and  \texttt{FormCalc}~\cite{Hahn:1998yk}.
We show the Feynman diagrams contributing to the real emission amplitude
and to the one-loop virtual correction in Fig.~\ref{fig:Hbb}.
We regularise IR  divergences which are present individually in both two- and
three-body contributions to the decay rate by performing loop integrals and phase-space integrals 
in $d=4-2\epsilon$ dimensions.  It is an important check on our calculation 
that these IR divergences cancel at the level of the decay rate, while UV divergences 
are removed by the counterterms.  A further check is that we reproduce the known SM 
results as a part of the full SMEFT calculation.   

We write the result for the NLO decay rate  in the SMEFT in the form
\begin{align}
\Gamma = \Gamma^{(4,0)}+\Gamma^{(4,1)}+ \Gamma^{(6,0)}+\Gamma^{(6,1)} \,,
\end{align}
where the first superscript differentiates between the SM and dimension-6 contributions, and the second
between powers of $\alpha_s$.  We provide results for the decay rate to $\mathcal{O}(\alpha_s)$
as an expansion in $\Lambda_{\rm NP}^{-2}$, 
keeping only the leading terms of $\mathcal{O}(\Lambda_{\rm NP}^{-2})$.  
The dimension-6 contributions to observables
thus appear through the interference of diagrams containing one dimension-6 operator with purely SM 
diagrams\footnote{Additional 
effects which may appear at $\mathcal{O}(\Lambda_{\rm NP}^{-4})$,
through interference of dimension-6 contributions or the introduction of dimension-8 operators
should be investigated if evidence for non-vanishing Wilson coefficients is observed at 
$\mathcal{O}(\Lambda_{\rm NP}^{-2})$. }.

In writing the results we shall make use of the shorthand notation 
\begin{align}
\beta = \sqrt{1-\frac{4 m_b^2}{m_h^2}} \,,\quad x = \frac{1 - \beta}{1 + \beta} \,,\quad y = \frac{1 - \beta^2}{4} 
=\frac{m_b^2}{m_h^2}  \,.
\end{align} 
Although the computation is performed with complex Wilson coefficients, 
the result for the decay rate depends only on the real parts of $C_{bH}$ and $C_{bG}$. 
To avoid cluttering the notation we do not write out, for instance, ${\rm Re} \,(C_{bH})$, but this is to be understood
in all the equations which follow.

The tree-level results for the decay rate are
\begin{align} \label{tree-level}
\nonumber
\Gamma^{(4,0)} &= \frac{N_c m_h m_b^2 \beta^3}{8 \pi v_T^2} \,, \\
\Gamma^{(6,0)} &= \left( 2 C_{H,{\rm kin}} - \frac{\sqrt{2} v_T^3}{m_b}C_{bH} \right) \Gamma^{(4,0)} \,. 
\end{align}
The QCD corrections for Higgs boson decays to massive quarks within the SM
have been known for a long time~\cite{Braaten:1980yq,Sakai:1980fa,Inami:1980qp,Drees:1989du,Drees:1990dq}.
The result is
\beq
\Gamma^{(4,1)} = \Gamma^{(4,0)} \frac{\alpha_s C_F}{\pi}  \frac{A(\beta)}{\beta^3}  \,,
\eeq
where the kinematic function $A(\beta)$ is given by
\begin{align}
\nonumber
A(\beta) & = 
	\frac{3\beta}{8} \left(-1+ 7 \beta^2\right) 
	+\beta^3 \left(3 \ln \left[y\right] - 4  \ln \left[\beta \right] \right)  \\ \nonumber
	&+ \ln \left[x\right] \bigg{\{}\frac{1}{16} (-3 - 34 \beta^2 + 13 \beta^4) \\ \nonumber
	&+ \beta^2 (1 + \beta^2) \left( -\frac{3}{2}\ln \left[y\right] + 2\ln \left[\beta\right]  \right) \bigg{\}} \\
    	&+ \beta^2 (1 + \beta^2) \left( \frac{3}{2} \ln^2 \left[x\right] + 2 {\rm Li}_2 \left[x\right] + {\rm Li}_2 \left[x^2\right] \right)\,.
\end{align}
The result for the $\mathcal{O}(\alpha_s)$ correction to the decay rate in the SMEFT
is a function of the Wilson coefficients $\{C_{bH},C_{H,{\rm kin}},C_{HG},C_{bG}\}$.
The first two appear at tree-level as in Eq.~(\ref{tree-level}), and the latter two contribute for the first time at $\mathcal{O}(\alpha_s)$.
The result is
\begin{align} \label{Anal}
\nonumber
\Gamma^{(6,1)} =& C_{bG} \frac{\alpha_s C_F}{\pi} \frac{N_c m_h^3 m_b}{8\sqrt{2}\pi v_T} \bigg{\{}
	 \frac{\beta}{8} \left(15 + 28 \beta^2 - 35 \beta^4\right)  \\ \nonumber
	&- \frac{3}{16} \left(-5 + 3 \beta^2 - 15 \beta^4 + 17 \beta^6\right) \ln \left[ x \right] \\ \nonumber
	&- 3 \beta^3 \left(1 - \beta^2\right) \ln \left[y \right]
\bigg{\}} \\[1mm]
\nonumber
+& C_{HG} \frac{\alpha_s C_F}{\pi} \frac{N_c m_h^2 m_b \sqrt{y}}{2 \pi}
\bigg{\{}
	\frac{\beta}{8} \left(15 - 2 \pi^2 \beta + 23 \beta^2\right) \\ \nonumber
	&-  \frac{3}{4} \beta^2 \ln^2 \left[ x \right]
	-  \frac{3}{2} \beta^3 \ln \left[y \right] \\ \nonumber
	&+  \ln \left[ x \right] \left(\frac{1}{16} (15 + 2 \beta^2 + 7 \beta^4) + \beta^2 \ln \left[ y \right] \right) \\ \nonumber
	&+ 3 \beta^2\left( {\rm Li}_2 \left[ x \right]- \frac{1}{2}  {\rm Li}_2 \left[ x^2 \right]\right)
\bigg{\}} \\[1mm]
%
\nonumber
+& 2\, \Gamma^{(4,1)} C_{H,{\rm kin}}	 \\ \nonumber
-& C_{bH} \frac{\alpha_s C_F}{\pi}  \frac{N_c m_h m_b v_T}{4\sqrt{2}\pi} \left( A(\beta) + \beta^3 - \frac{3}{4}\beta^3 \ln \left[ y \right] \right) 
 \\[1mm]
 +& \Gamma^{(4,0)} \frac{v_T^3}{\sqrt{2} m_b} \frac{\dot{C}_{bH}}{(4\pi)^2}  \ln \left[\frac{\mu^2}{m_H^2} \right] \, ,
\end{align}
where the $\mathcal{O}(\alpha_s)$ contributions to $\dot{C}_{bH}$  can be deduced from Eq.~(\ref{delCbh}). 

The expressions above are a main result of this work. We have
written the results such that the explicit powers of $m_b$ 
correspond to those generated by the Yukawa coupling appearing in
$hb\bar b$-vertex, while implicit powers (through, e.g., $y$) are generated through phase-space
factors or through propagators.  This will prove convenient when discussing
results in the \msbar~scheme for the $b$-quark mass in the next section.
We note that the QCD corrections involving $C_{bH}$ do not factorise explicitly 
in the above expression, that is, they are not proportional to the 
SM ones. This is a consequence of renormalising the $b$-quark mass and 
the Wilson coefficients in different schemes, as we shall see below.
\subsection{Results in the \msbar~scheme and the massless limit}
The results presented in the previous section are valid
in the on-shell scheme for the masses, and the \msbar~scheme for the Wilson
coefficients. In the presence of a large separation between the scales $m_b$
and $m_h$, the fixed-order predictions become inappropriate for phenomenology due the
appearance of large logarithms of $m_h/m_b$ which deteriorate the convergence of the 
perturbative series in $\alpha_s$. 

To overcome this, the decay rate predictions can be converted
into the \msbar~scheme for the $b$-quark mass (we 
refer to such predictions as being in the \msbar~scheme for the decay rate).
In this scheme, the dominant large logarithmic corrections
are resummed into RG evolution factors relating the value of the running 
$b$-quark mass  at different energy scales.
Mass renormalisation in the \msbar~scheme is achieved by dropping the finite 
contributions to the $b$-quark mass counterterm in Eq.~(\ref{eq:4fctsB}). 
This implies that the relation between the $b$-quark pole mass and \msbar~mass is
\begin{align}
\label{eq:mbConv}
m_b &= \overline{m}_b(\mu) \left(1 - \delta_m (\mu) \right) \,,
\end{align}
where the SM and dimension-6 contributions to $\delta_m(\mu)$ are
\begin{align} \label{poleConv}
\nonumber
\delta_m^{(4)}(\mu) &= -\frac{\alpha_s C_F}{\pi}   \left(1 +\frac{3}{4} \ln \left[ \frac{\mu^2}{\overline{m}_b^2} \right] \right)\,,\\
\delta_m^{(6)}(\mu) &= -\frac{\alpha_s C_F}{\pi} \frac{v_T \overline{m}_b}{\sqrt{2}} C_{bG} \left(1 + 3 \ln \left[ \frac{\mu^2}{\overline{m}_b^2} \right] \right)\,.
\end{align}
We can then define tree-level results in the \msbar~scheme as\footnote{Note that we have not used the \msbar~mass in the $\beta^3$ terms, which are related to phase-space factors for on-shell quark production rather than
the Yukawa coupling of the Higgs.  If the $m_b$ appearing in $\beta$ were also converted, one 
would need to add extra terms to Eq.~(\ref{MSbar}).  However, we have checked that the numerical 
results obtained in that way are nearly identical to those given below, so we have opted for the more streamlined
(and physically motivated)  version where the phase-space factor is kept in the on-shell scheme.}
\begin{align} 
\nonumber
\overline{\Gamma}^{(4,0)} =& \frac{N_c m_h \overline{m}_b^2 \beta^3}{8 \pi v_T^2} \, ,\\ 
\overline{\Gamma}^{(6,0)} =& \left(2 C_{H, \rm kin} - \frac{\sqrt{2} v_T^3}{\overline{m}_b}C_{bH} \right)\overline{\Gamma}^{(4,0)} \, .
\end{align}
The corresponding results for the decay rate in the \msbar~scheme are therefore
\begin{align} \label{MSbar}
\nonumber
\overline{\Gamma}^{(4,1)} &= \Gamma^{(4,1)} - 2 \delta_m^{(4)}(\mu) \overline{\Gamma}^{(4,0)} \,,\\ \nonumber
\overline{\Gamma}^{(6,1)} &= \Gamma^{(6,1)} - 2 \delta_m^{(6)}(\mu) \overline{\Gamma}^{(4,0)}  \\
	&\phantom{xx}-2 \delta_m^{(4)}(\mu) \left( 2 C_{H,{\rm kin}} - \frac{v_T^3}{\sqrt{2} \overline{m}_b}C_{bH} \right) \overline{\Gamma}^{(4,0)}\,,
\end{align}
where it is understood that $m_b\to \overline{m}_b$ in the first term on the right-hand side of each of the above equations.
To obtain the leading-logarithmic (LL) solution for the running mass, both
SM and dimension-6 corrections to the $b$-quark mass must be taken into
account. By inspecting Eq.~(\ref{eq:mbConv}), 
the following differential equation should be solved
\begin{align} \label{RunMass}
\frac{d\,\overline{m}_b}{d \ln(\mu)}  &= - \frac{\alpha_s C_F}{\pi} \frac{3}{2}\overline{m}_b\left(1 + 2\sqrt{2}v_T \overline{m}_b C_{bG} \right)+\mathcal{O}(\alpha_s^2) \,.
\end{align}
This can be achieved analytically by first finding the LL solution for the running of $C_{bG}(\mu)$,
and in addition the running $b$-quark mass in the SM --- which we label as $\overline{m}_{b}^{(4)}$. 
In doing so we adopt the following convention for the QCD $\beta$-function
\begin{align}\label{betaFn}
\frac{d }{d \ln(\mu)}\frac{\alpha_s(\mu)}{\pi} &= - 2 \beta_0 \left( \frac{\alpha_s(\mu)}{\pi} \right)^2\,+...,
\end{align}
where the ellipses refer to higher-order terms in $\alpha_s$ and $\Lambda_{\rm NP}^{-2}$, and
$\beta_0 = (11 N_c-2 n_f)/12$, with $n_f = 5$ the number of active flavours.
A solution for $C_{bG}(\mu)$ can easily be obtained when taking into account
only the numerically important self-mixing contribution~\cite{Alonso:2013hga}. 
This is achieved by solving the equation
\begin{align}
\frac{d C_{bG}(\mu)}{d \ln(\mu)}  = - 2 \frac{\alpha_s(\mu)}{\pi} \gamma_C^0 C_{bG}(\mu) \,.
\end{align}
We find
\begin{align}\label{CbGRun}
C_{bG}(\mu) &= C_{bG}(\mu_0)  \left( \frac{\alpha_s(\mu)}{\alpha_s(\mu_0)}\right)^{\frac{\gamma_C^0}{\beta_0}} \,, \phantom{x} \gamma_C^0 = -\frac{5 C_F -2 N_c}{4} \, .
\end{align}
Note that our result for $\gamma_C^0$ differs with respect to that presented in~\cite{Alonso:2013hga},
since our operator definition in Table~\ref{op59} implies $C_{bG} = \overline{C}_{bG}/g_s$,  where $\overline{C}_{bG}$ is the definition used in ~\cite{Alonso:2013hga}.
Writing
\begin{align}\label{bmassD4}
\overline{m}_{b}^{(4)}(\mu) &= \overline{m}_{b}(\mu_0) \left( \frac{\alpha_s(\mu)}{\alpha_s(\mu_0)}\right)^{\frac{\gamma_m^0}{\beta_0}} \,,  \phantom{x} \gamma_m^0 = \frac{3}{4} C_F \,,
\end{align}
a solution for $\overline{m}_{b}$ to $\mathcal{O}(\Lambda_{\rm NP}^{-2})$ can then be obtained as
\begin{align} \label{bmassQCD}
\nonumber
\overline{m}_{b}(\mu) &= \overline{m}_{b}(\mu_0) \left( \frac{\alpha_s(\mu)}{\alpha_s(\mu_0)}\right)^{\frac{\gamma_m^0}{\beta_0}}
\bigg(1 +\frac{2\sqrt{2}v_T}{\gamma_m^0+\gamma_c^0}\\
	&\phantom{xx}\left[ \overline{m}_{b}^{(4)}(\mu)C_{bG}(\mu) -  \overline{m}_{b}^{(4)}(\mu_0)C_{bG}(\mu_0)\right] \bigg)  \,.
\end{align}

The motivation for using the \msbar~scheme for the $b$-quark
mass is to resum large logarithms introduced by the scale ratio
$m_b/m_h\ll 1$. 
Under such circumstances, it also makes sense to consider the
massless limit of the decay rate, obtained as $\beta \to 1$, which as we will
show below is an excellent approximation numerically.  Keeping the first non-vanishing terms
in the SM and for each Wilson coefficient in the SMEFT, we find 
\begin{align} \label{Massless}
\nonumber
\overline{\Gamma}^{(4,0)}_{\beta\to1} =& \frac{N_c m_h \overline{m}_b^2}{8 \pi v_T^2} \,, \\[1mm] \nonumber
\overline{\Gamma}^{(4,1)}_{\beta\to1} =& \frac{\alpha_s C_F}{\pi} \frac{1}{4} \left( 17 + 6 \ln\left[\frac{\mu^2}{m_h^2}\right] \right)\overline{\Gamma}^{(4,0)}_{\beta\to1}\,, \\[1mm] \nonumber
\overline{\Gamma}^{(6,0)}_{\beta\to1} =& \left(2 C_{H, \rm kin} - \frac{\sqrt{2} v_T^3}{\overline{m}_b}C_{bH} \right)\overline{\Gamma}^{(4,0)}_{\beta \to1} \,, \\[1mm] \nonumber
\overline{\Gamma}^{(6,1)}_{\beta\to1} =& \left( 2 C_{H,{\rm kin}} - \frac{\sqrt{2} v_T^3}{\overline{m}_b} C_{bH} \right) \overline{\Gamma}^{(4,1)}_{\beta\to1} \\ \nonumber
	&+ \frac{\alpha_s C_F}{\pi} \frac{N_c m_h^3 \overline{m}_b}{8\sqrt{2} \pi v_T} C_{bG} 
	+ \frac{\alpha_s C_F}{\pi} \frac{N_c m_h \overline{m}_b^2}{8 \pi} C_{HG} \\ 
	& \times \bigg(19-\pi^2 +\ln^2\left[\frac{\overline{m}_b^2}{m_h^2}\right] +6 \ln\left[\frac{\mu^2}{m_h^2}\right] \bigg) \,,
\end{align}
where we have used 
\begin{equation}
A(\beta \to 1) = \frac{9}{4}+ \frac{3}{2}\ln\left[ \frac{m_b^2}{m_h^2}\right] \,.
\end{equation}
Furthermore, in the massless limit the dimension-6 corrections from $C_{bG}$ to the 
running mass  in Eq.~(\ref{bmassQCD}) vanish.  It is worth mentioning that even though
the phase-space factor multiplying  $C_{HG}$ vanishes as $m_b\to 0$ (because of the 
factor of $\sqrt{y}$ in Eq.~(\ref{Anal})), it is multiplied by a large logarithm in the ratio of $m_b/m_h$,
which is not removed by the conversion to the \msbar~scheme.  Finally, in the \msbar~scheme the coefficients
$C_{bH}$ and $C_{H, {\rm kin}}$, which first appear at tree level, receive NLO QCD corrections
proportional to the SM ones, while the coefficients $C_{HG}$ and $C_{bG}$ do not.

To test the validity of the massless approximation, we
compare numerically the predictions for the total decay rate in the massless
limit with the full result. Specifically, we compare
\begin{align}
\overline{\Gamma} = \overline{\Gamma}^{(4,0)} + \overline{\Gamma}^{(4,1)} + \overline{\Gamma}^{(6,0)} + \overline{\Gamma}^{(6,1)} \,,
\qquad \overline{\Gamma}_{\beta\to1} \,.
\end{align}
The following set of inputs are used: $\alpha_s\left(m_Z\right) = 0.1184$, 
$m_Z = 91.1876$~GeV, $\overline{m}_b(\overline{m}_b) = 4.18$~GeV, $m_t = 173.0$~GeV and $m_h = 125.0$~GeV.
The replacement 
$v_T\to(\sqrt{2} G_F)^{-\frac{1}{2}}$ is made with $G_F = 1.16637\cdot10^{-5}$~GeV$^{-2}$ for numerical evaluation, though the results are presented with an explicit factor of $v_T\approx 246$~GeV in order compare their size relative to the electroweak scale. 
We also introduce the dimensionless Wilson coefficients $\tilde{C}=\Lambda_{\rm NP}^2 C$, except for $C_{H, \rm kin}$ where we also extract the factor of $v_T^2$ and write $\tilde{C}_{H, \rm kin}=\left(\Lambda_{\rm NP}/v_T\right)^2 C_{H, \rm kin}$.
At the scale $\mu = m_h$, and using Eqs.~(\ref{betaFn}), (\ref{CbGRun}) and (\ref{bmassQCD}) to run the various parameters, we obtain
\begin{align}
\nonumber
\frac{\overline{\Gamma}}{\rm MeV}  &= \kappa^{\rm QCD} \bigg\{ 2.22 \left[1 +  2 \left(\frac{v_T}{\Lambda_{\rm NP}}\right)^2 \tilde{C}_{H,{\rm kin}}\right]  \\  \nonumber
& \phantom{xxxxxxxx} - 258 \left(\frac{v_T}{\Lambda_{\rm NP}}\right)^2 \tilde{C}_{bH} \bigg\} 
   \\ \nonumber
	&\phantom{x}+\left(\frac{v_T}{\Lambda_{\rm NP}}\right)^2(1.55 \tilde{C}_{bG} + 6.88 \tilde{C}_{HG})+...\,,\\[1mm]
\nonumber
\frac{\overline{\Gamma}_{\beta\to1}}{\rm MeV}  &= \kappa^{\rm QCD}_{\beta\to1} \bigg\{ 2.23 \left[1 +  2 \left(\frac{v_T}{\Lambda_{\rm NP}}\right)^2 \tilde{C}_{H,{\rm kin}}\right] \\ \nonumber
& \phantom{xxxxxxxx}  - 257\left(\frac{v_T}{\Lambda_{\rm NP}}\right)^2 \tilde{C}_{bH} \bigg\} \\
	&\phantom{x}+ \left(\frac{v_T}{\Lambda_{\rm NP}}\right)^2 \left(1.57 \tilde{C}_{bG}+6.91 \tilde{C}_{HG}\right) \,,
\end{align}
where we have introduced $\kappa^{\rm QCD} \approx \kappa^{\rm QCD}_{\beta\to1} \approx 1.20$ to highlight the impact of the QCD corrections on the tree-level result.
The ellipses denote terms of $\mathcal{O}(\Lambda_{\rm NP}^{-4})$ which are generated
by the running $b$-quark mass (see Eq.~\ref{bmassQCD}) and are higher-order in the power counting.
The massless limit is therefore found to be an extremely good approximation.

We end this section by commenting on the possible impact of our NLO
calculation on global fits to Higgs data.  As a concrete example,
consider the scenario where a future experiment observes a 5\%
deviation in the partial width $\Gamma_{h\to b\bar b}$ compared to its
SM value.  Under the assumption that the Wilson coefficients appearing
in Eq.~(\ref{Massless}) are $\mathcal{O}(1)$, the contributions
involving $\tilde{C}_{HG}$, $\tilde{C}_{bG}$, $\tilde{C}_{H,{\rm
    kin}}$ can be ignored and only the Wilson coefficient
$\tilde{C}_{bH}$ is relevant.  The NLO corrections increase the
sensitivity to this coefficient by about 20\% compared to the LO
calculation, and such a measurement on the partial width could be used
to probe scales $\Lambda_{\rm NP} \approx 10$~TeV.  However, as
discussed in~\cite{Elias-Miro:2013mua}, in a broad range of UV
complete models it is expected that the Wilson coefficient
$\tilde{C}_{bH}$ scales as $\tilde{C}^{\rm MFV}_{bH}\sim y_b
\tilde{C}_{bH}$.  In that case, even though the prefactor multiplying
$\tilde{C}_{HG}$ is about 45 times smaller than that multiplying
$\tilde{C}_{bH}$, the sensitivity to $\tilde{C}^{\rm MFV}_{bH}$,
$\tilde{C}_{HG}$, and all the other Wilson coefficients is roughly of
the same order since $45 y_b \sim 1$, and scales $\Lambda_{\rm NP}
\approx 2$~TeV would be probed.  A purely LO calculation would miss
the contributions from $\tilde{C}_{HG}$ and $\tilde{C}_{bG}$ entirely.



\section{Discussion and conclusions}
As discussed in the introduction, the Higgs boson couplings are inferred
by performing a (global) fit to $\sigma \cdot BR(H\to X)$ 
data of the observable final states.
Many groups have now performed dedicated analyses and/or global fits to the 
LHC Higgs coupling data in an EFT framework~\cite{Elias-Miro:2013gya,Azatov:2012bz,Espinosa:2012im,Plehn:2012iz,Carmi:2012in,Peskin:2012we,Corbett:2012ja,Masso:2012eq,Grojean:2013kd,Falkowski:2013dza,Dumont:2013wma,Djouadi:2013qya,Elias-Miro:2013mua,Lopez-Val:2013yba,Pomarol:2013zra,Elias-Miro:2013eta,Englert:2014uua,Ellis:2014dva,Belusca-Maito:2014dpa,Ellis:2014jta,Hartmann:2015oia,Corbett:2015ksa,Hartmann:2015aia,Buchalla:2015qju,Englert:2015hrx,Aad:2015tna,Butter:2016cvz,Cirigliano:2016nyn}.
It should be noted that including constraints from precision electroweak and low-energy observables
is also important in a global fit to Higgs data at NLO, since operators which contribute to electroweak
precision observables at tree-level can enter the expressions for NLO Higgs decay rates~\cite{Chen:2013kfa}.

An important tool used in several of these global fits is the package \texttt{eHDECAY}~\cite{Contino:2013kra,Contino:2014aaa} 
which allows the computation of Higgs boson decay widths and branching ratios within the SMEFT (implemented
in the SILH basis~\cite{Giudice:2007fh}). This program is an extension of \texttt{HDECAY}~\cite{Djouadi:1997yw,Butterworth:2010ym}, which computes these observables within the SM\footnote{Numerical values for a range of inputs have been
provided by the LHC Higgs Cross Section Working Group~\cite{Heinemeyer:2013tqa}.}.
In this case, the expression for the $h\to b\bar b$ EFT decay rate is obtained by applying 
a scaling of the tree-level prediction. The scaling factor is the same as that used in the SM,
and includes the massless QCD corrections up to $\mathcal{O}(\alpha_s^4)$~\cite{Gorishnii:1983cu,Gorishnii:1991zr,Kataev:1993be,Surguladze:1994gc,Larin:1995sq,Chetyrkin:1995pd,Chetyrkin:1996sr,Chetyrkin:1997vj,Baikov:2005rw}.
This is of course a reasonable procedure, since the QCD corrections to the $hb\bar b$-vertex 
involving $C_{bH}$ and $C_{H,{\rm kin}}$ factorise, and the most accurate predictions should be 
applied wherever possible.
In addition to the scaled tree-level results, we suggest to include the $\mathcal{O}(\alpha_s)$ 
contributions involving $C_{bG}$ and $C_{HG}$ which are currently not included. 
The relevant results are provided in Eq.~(\ref{Massless}).
It should be noted that the dominant contribution from both of these operators 
arise from genuine NLO effects, which cannot be predicted by an RG analysis.

Finally, electroweak corrections are known for the $h\to b\bar b$ decay rate in the SM
~\cite{Fleischer:1980ub,Bardin:1990zj,Dabelstein:1991ky,Kniehl:1991ze,Mihaila:2015lwa}. However, unlike 
the QCD corrections involving $C_{bH}$ and $C_{H,{\rm kin}}$, these contributions do not factorise 
in a straightforward way, and a scaling by known SM corrections is not appropriate. 
A dedicated calculation of the full NLO electroweak corrections in the SMEFT is in progress. 
In the meantime, we suggest to include the results involving the one-loop 
four-fermion and dominant Yukawa corrections which were computed in~\cite{Gauld:2015lmb}.
In this case, the running of the $b$-quark mass receives large logarithmic
corrections involving the scalar four-fermion operators $Q_{qtqb}$, and these contributions
should be resummed following a similar procedure as was taken for $Q_{bG}$ in this work.

\section{Acknowledgements}
The research of D. J. S. is supported by an STFC Postgraduate Studentship.

\bibliography{RG}

\begin{thebibliography}{108}
\expandafter\ifx\csname natexlab\endcsname\relax\def\natexlab#1{#1}\fi
\expandafter\ifx\csname bibnamefont\endcsname\relax
  \def\bibnamefont#1{#1}\fi
\expandafter\ifx\csname bibfnamefont\endcsname\relax
  \def\bibfnamefont#1{#1}\fi
\expandafter\ifx\csname citenamefont\endcsname\relax
  \def\citenamefont#1{#1}\fi
\expandafter\ifx\csname url\endcsname\relax
  \def\url#1{\texttt{#1}}\fi
\expandafter\ifx\csname urlprefix\endcsname\relax\def\urlprefix{URL }\fi
\providecommand{\bibinfo}[2]{#2}
\providecommand{\eprint}[2][]{\url{#2}}

\bibitem[{\citenamefont{Aad et~al.}(2012)}]{Aad:2012tfa}
\bibinfo{author}{\bibfnamefont{G.}~\bibnamefont{Aad}} \bibnamefont{et~al.}
  (\bibinfo{collaboration}{ATLAS Collaboration}), \bibinfo{journal}{Phys.Lett.}
  \textbf{\bibinfo{volume}{B716}}, \bibinfo{pages}{1} (\bibinfo{year}{2012}),
  \eprint{1207.7214}.

\bibitem[{\citenamefont{Chatrchyan et~al.}(2012)}]{Chatrchyan:2012ufa}
\bibinfo{author}{\bibfnamefont{S.}~\bibnamefont{Chatrchyan}}
  \bibnamefont{et~al.} (\bibinfo{collaboration}{CMS Collaboration}),
  \bibinfo{journal}{Phys.Lett.} \textbf{\bibinfo{volume}{B716}},
  \bibinfo{pages}{30} (\bibinfo{year}{2012}), \eprint{1207.7235}.

\bibitem[{\citenamefont{Bolognesi et~al.}(2012)\citenamefont{Bolognesi, Gao,
  Gritsan, Melnikov, Schulze, Tran, and Whitbeck}}]{Bolognesi:2012mm}
\bibinfo{author}{\bibfnamefont{S.}~\bibnamefont{Bolognesi}},
  \bibinfo{author}{\bibfnamefont{Y.}~\bibnamefont{Gao}},
  \bibinfo{author}{\bibfnamefont{A.~V.} \bibnamefont{Gritsan}},
  \bibinfo{author}{\bibfnamefont{K.}~\bibnamefont{Melnikov}},
  \bibinfo{author}{\bibfnamefont{M.}~\bibnamefont{Schulze}},
  \bibinfo{author}{\bibfnamefont{N.~V.} \bibnamefont{Tran}}, \bibnamefont{and}
  \bibinfo{author}{\bibfnamefont{A.}~\bibnamefont{Whitbeck}},
  \bibinfo{journal}{Phys. Rev.} \textbf{\bibinfo{volume}{D86}},
  \bibinfo{pages}{095031} (\bibinfo{year}{2012}), \eprint{1208.4018}.

\bibitem[{\citenamefont{Aad et~al.}(2013)}]{Aad:2013xqa}
\bibinfo{author}{\bibfnamefont{G.}~\bibnamefont{Aad}} \bibnamefont{et~al.}
  (\bibinfo{collaboration}{ATLAS}), \bibinfo{journal}{Phys. Lett.}
  \textbf{\bibinfo{volume}{B726}}, \bibinfo{pages}{120} (\bibinfo{year}{2013}),
  \eprint{1307.1432}.

\bibitem[{\citenamefont{Aad et~al.}(2015{\natexlab{a}})}]{Aad:2015gba}
\bibinfo{author}{\bibfnamefont{G.}~\bibnamefont{Aad}} \bibnamefont{et~al.}
  (\bibinfo{collaboration}{ATLAS}) (\bibinfo{year}{2015}{\natexlab{a}}),
  \eprint{1507.04548}.

\bibitem[{\citenamefont{Khachatryan et~al.}(2015)}]{Khachatryan:2014jba}
\bibinfo{author}{\bibfnamefont{V.}~\bibnamefont{Khachatryan}}
  \bibnamefont{et~al.} (\bibinfo{collaboration}{CMS}), \bibinfo{journal}{Eur.
  Phys. J.} \textbf{\bibinfo{volume}{C75}}, \bibinfo{pages}{212}
  (\bibinfo{year}{2015}), \eprint{1412.8662}.

\bibitem[{ATL(2015)}]{ATLAS-CONF-2015-044}
\bibinfo{type}{Tech. Rep.} \bibinfo{number}{ATLAS-CONF-2015-044},
  \bibinfo{institution}{CERN}, \bibinfo{address}{Geneva}
  (\bibinfo{year}{2015}), \urlprefix\url{http://cds.cern.ch/record/2052552}.

\bibitem[{\citenamefont{Dawson et~al.}(2013)}]{Dawson:2013bba}
\bibinfo{author}{\bibfnamefont{S.}~\bibnamefont{Dawson}} \bibnamefont{et~al.}
  (\bibinfo{year}{2013}), \eprint{1310.8361},
  \urlprefix\url{http://inspirehep.net/record/1262795/files/arXiv:1310.8361.pdf}.

\bibitem[{\citenamefont{Andersen et~al.}(2013)}]{Heinemeyer:2013tqa}
\bibinfo{author}{\bibfnamefont{J.~R.} \bibnamefont{Andersen}}
  \bibnamefont{et~al.} (\bibinfo{collaboration}{LHC Higgs Cross Section Working
  Group}) (\bibinfo{year}{2013}), \eprint{1307.1347}.

\bibitem[{\citenamefont{Chatrchyan et~al.}(2014)}]{Chatrchyan:2013zna}
\bibinfo{author}{\bibfnamefont{S.}~\bibnamefont{Chatrchyan}}
  \bibnamefont{et~al.} (\bibinfo{collaboration}{CMS}), \bibinfo{journal}{Phys.
  Rev.} \textbf{\bibinfo{volume}{D89}}, \bibinfo{pages}{012003}
  (\bibinfo{year}{2014}), \eprint{1310.3687}.

\bibitem[{\citenamefont{Aad et~al.}(2015{\natexlab{b}})}]{Aad:2014xzb}
\bibinfo{author}{\bibfnamefont{G.}~\bibnamefont{Aad}} \bibnamefont{et~al.}
  (\bibinfo{collaboration}{ATLAS}), \bibinfo{journal}{JHEP}
  \textbf{\bibinfo{volume}{01}}, \bibinfo{pages}{069}
  (\bibinfo{year}{2015}{\natexlab{b}}), \eprint{1409.6212}.

\bibitem[{\citenamefont{Dixon and Li}(2013)}]{Dixon:2013haa}
\bibinfo{author}{\bibfnamefont{L.~J.} \bibnamefont{Dixon}} \bibnamefont{and}
  \bibinfo{author}{\bibfnamefont{Y.}~\bibnamefont{Li}}, \bibinfo{journal}{Phys.
  Rev. Lett.} \textbf{\bibinfo{volume}{111}}, \bibinfo{pages}{111802}
  (\bibinfo{year}{2013}), \eprint{1305.3854}.

\bibitem[{\citenamefont{Caola and Melnikov}(2013)}]{Caola:2013yja}
\bibinfo{author}{\bibfnamefont{F.}~\bibnamefont{Caola}} \bibnamefont{and}
  \bibinfo{author}{\bibfnamefont{K.}~\bibnamefont{Melnikov}},
  \bibinfo{journal}{Phys. Rev.} \textbf{\bibinfo{volume}{D88}},
  \bibinfo{pages}{054024} (\bibinfo{year}{2013}), \eprint{1307.4935}.

\bibitem[{\citenamefont{Campbell
  et~al.}(2014{\natexlab{a}})\citenamefont{Campbell, Ellis, and
  Williams}}]{Campbell:2013una}
\bibinfo{author}{\bibfnamefont{J.~M.} \bibnamefont{Campbell}},
  \bibinfo{author}{\bibfnamefont{R.~K.} \bibnamefont{Ellis}}, \bibnamefont{and}
  \bibinfo{author}{\bibfnamefont{C.}~\bibnamefont{Williams}},
  \bibinfo{journal}{JHEP} \textbf{\bibinfo{volume}{04}}, \bibinfo{pages}{060}
  (\bibinfo{year}{2014}{\natexlab{a}}), \eprint{1311.3589}.

\bibitem[{\citenamefont{Campbell
  et~al.}(2014{\natexlab{b}})\citenamefont{Campbell, Ellis, and
  Williams}}]{Campbell:2013wga}
\bibinfo{author}{\bibfnamefont{J.~M.} \bibnamefont{Campbell}},
  \bibinfo{author}{\bibfnamefont{R.~K.} \bibnamefont{Ellis}}, \bibnamefont{and}
  \bibinfo{author}{\bibfnamefont{C.}~\bibnamefont{Williams}},
  \bibinfo{journal}{Phys. Rev.} \textbf{\bibinfo{volume}{D89}},
  \bibinfo{pages}{053011} (\bibinfo{year}{2014}{\natexlab{b}}),
  \eprint{1312.1628}.

\bibitem[{\citenamefont{Campbell
  et~al.}(2014{\natexlab{c}})\citenamefont{Campbell, Ellis, Furlan, and
  Rontsch}}]{Campbell:2014gua}
\bibinfo{author}{\bibfnamefont{J.~M.} \bibnamefont{Campbell}},
  \bibinfo{author}{\bibfnamefont{R.~K.} \bibnamefont{Ellis}},
  \bibinfo{author}{\bibfnamefont{E.}~\bibnamefont{Furlan}}, \bibnamefont{and}
  \bibinfo{author}{\bibfnamefont{R.}~\bibnamefont{Rontsch}},
  \bibinfo{journal}{Phys. Rev.} \textbf{\bibinfo{volume}{D90}},
  \bibinfo{pages}{093008} (\bibinfo{year}{2014}{\natexlab{c}}),
  \eprint{1409.1897}.

\bibitem[{\citenamefont{Englert et~al.}(2016)\citenamefont{Englert, McCullough,
  and Spannowsky}}]{Englert:2015bwa}
\bibinfo{author}{\bibfnamefont{C.}~\bibnamefont{Englert}},
  \bibinfo{author}{\bibfnamefont{M.}~\bibnamefont{McCullough}},
  \bibnamefont{and}
  \bibinfo{author}{\bibfnamefont{M.}~\bibnamefont{Spannowsky}},
  \bibinfo{journal}{Nucl. Phys.} \textbf{\bibinfo{volume}{B902}},
  \bibinfo{pages}{440} (\bibinfo{year}{2016}), \eprint{1504.02458}.

\bibitem[{\citenamefont{Khachatryan et~al.}(2014)}]{Khachatryan:2014iha}
\bibinfo{author}{\bibfnamefont{V.}~\bibnamefont{Khachatryan}}
  \bibnamefont{et~al.} (\bibinfo{collaboration}{CMS}), \bibinfo{journal}{Phys.
  Lett.} \textbf{\bibinfo{volume}{B736}}, \bibinfo{pages}{64}
  (\bibinfo{year}{2014}), \eprint{1405.3455}.

\bibitem[{\citenamefont{Aad et~al.}(2015{\natexlab{c}})}]{Aad:2015xua}
\bibinfo{author}{\bibfnamefont{G.}~\bibnamefont{Aad}} \bibnamefont{et~al.}
  (\bibinfo{collaboration}{ATLAS}), \bibinfo{journal}{Eur. Phys. J.}
  \textbf{\bibinfo{volume}{C75}}, \bibinfo{pages}{335}
  (\bibinfo{year}{2015}{\natexlab{c}}), \eprint{1503.01060}.

\bibitem[{\citenamefont{Aicheler et~al.}(2012)\citenamefont{Aicheler, Aicheler,
  Burrows, Draper, Garvey, Lebrun, Peach, Phinney, Schmickler, Schulte
  et~al.}}]{Aicheler:2012bya}
\bibinfo{author}{\bibfnamefont{M.}~\bibnamefont{Aicheler}},
  \bibinfo{author}{\bibfnamefont{M.}~\bibnamefont{Aicheler}},
  \bibinfo{author}{\bibfnamefont{P.}~\bibnamefont{Burrows}},
  \bibinfo{author}{\bibfnamefont{M.}~\bibnamefont{Draper}},
  \bibinfo{author}{\bibfnamefont{T.}~\bibnamefont{Garvey}},
  \bibinfo{author}{\bibfnamefont{P.}~\bibnamefont{Lebrun}},
  \bibinfo{author}{\bibfnamefont{K.}~\bibnamefont{Peach}},
  \bibinfo{author}{\bibfnamefont{N.}~\bibnamefont{Phinney}},
  \bibinfo{author}{\bibfnamefont{H.}~\bibnamefont{Schmickler}},
  \bibinfo{author}{\bibfnamefont{D.}~\bibnamefont{Schulte}},
  \bibnamefont{et~al.} (\bibinfo{year}{2012}).

\bibitem[{\citenamefont{Behnke et~al.}(2013)\citenamefont{Behnke, Brau, Foster,
  Fuster, Harrison, Paterson, Peskin, Stanitzki, Walker, and
  Yamamoto}}]{Behnke:2013xla}
\bibinfo{author}{\bibfnamefont{T.}~\bibnamefont{Behnke}},
  \bibinfo{author}{\bibfnamefont{J.~E.} \bibnamefont{Brau}},
  \bibinfo{author}{\bibfnamefont{B.}~\bibnamefont{Foster}},
  \bibinfo{author}{\bibfnamefont{J.}~\bibnamefont{Fuster}},
  \bibinfo{author}{\bibfnamefont{M.}~\bibnamefont{Harrison}},
  \bibinfo{author}{\bibfnamefont{J.~M.} \bibnamefont{Paterson}},
  \bibinfo{author}{\bibfnamefont{M.}~\bibnamefont{Peskin}},
  \bibinfo{author}{\bibfnamefont{M.}~\bibnamefont{Stanitzki}},
  \bibinfo{author}{\bibfnamefont{N.}~\bibnamefont{Walker}}, \bibnamefont{and}
  \bibinfo{author}{\bibfnamefont{H.}~\bibnamefont{Yamamoto}}
  (\bibinfo{year}{2013}), \eprint{1306.6327}.

\bibitem[{\citenamefont{Alexahin et~al.}(2013)}]{Alexahin:2013ojp}
\bibinfo{author}{\bibfnamefont{Y.}~\bibnamefont{Alexahin}} \bibnamefont{et~al.}
  (\bibinfo{year}{2013}), \eprint{1308.2143},
  \urlprefix\url{https://inspirehep.net/record/1247266/files/arXiv:1308.2143.pdf}.

\bibitem[{\citenamefont{Delahaye et~al.}(2013)}]{Delahaye:2013jla}
\bibinfo{author}{\bibfnamefont{J.-P.} \bibnamefont{Delahaye}}
  \bibnamefont{et~al.} (\bibinfo{year}{2013}), \eprint{1308.0494},
  \urlprefix\url{https://inspirehep.net/record/1246163/files/arXiv:1308.0494.pdf}.

\bibitem[{\citenamefont{Asner et~al.}(2003)\citenamefont{Asner, Gronberg, and
  Gunion}}]{Asner:2001ia}
\bibinfo{author}{\bibfnamefont{D.~M.} \bibnamefont{Asner}},
  \bibinfo{author}{\bibfnamefont{J.~B.} \bibnamefont{Gronberg}},
  \bibnamefont{and} \bibinfo{author}{\bibfnamefont{J.~F.}
  \bibnamefont{Gunion}}, \bibinfo{journal}{Phys. Rev.}
  \textbf{\bibinfo{volume}{D67}}, \bibinfo{pages}{035009}
  (\bibinfo{year}{2003}), \eprint{hep-ph/0110320}.

\bibitem[{\citenamefont{Bogacz et~al.}(2012)\citenamefont{Bogacz, Ellis,
  Lusito, Schulte, Takahashi, Velasco, Zanetti, and
  Zimmermann}}]{Bogacz:2012fs}
\bibinfo{author}{\bibfnamefont{S.~A.} \bibnamefont{Bogacz}},
  \bibinfo{author}{\bibfnamefont{J.}~\bibnamefont{Ellis}},
  \bibinfo{author}{\bibfnamefont{L.}~\bibnamefont{Lusito}},
  \bibinfo{author}{\bibfnamefont{D.}~\bibnamefont{Schulte}},
  \bibinfo{author}{\bibfnamefont{T.}~\bibnamefont{Takahashi}},
  \bibinfo{author}{\bibfnamefont{M.}~\bibnamefont{Velasco}},
  \bibinfo{author}{\bibfnamefont{M.}~\bibnamefont{Zanetti}}, \bibnamefont{and}
  \bibinfo{author}{\bibfnamefont{F.}~\bibnamefont{Zimmermann}}
  (\bibinfo{year}{2012}), \eprint{1208.2827}.

\bibitem[{\citenamefont{Baer et~al.}(2013)\citenamefont{Baer, Barklow, Fujii,
  Gao, Hoang, Kanemura, List, Logan, Nomerotski, Perelstein
  et~al.}}]{Baer:2013cma}
\bibinfo{author}{\bibfnamefont{H.}~\bibnamefont{Baer}},
  \bibinfo{author}{\bibfnamefont{T.}~\bibnamefont{Barklow}},
  \bibinfo{author}{\bibfnamefont{K.}~\bibnamefont{Fujii}},
  \bibinfo{author}{\bibfnamefont{Y.}~\bibnamefont{Gao}},
  \bibinfo{author}{\bibfnamefont{A.}~\bibnamefont{Hoang}},
  \bibinfo{author}{\bibfnamefont{S.}~\bibnamefont{Kanemura}},
  \bibinfo{author}{\bibfnamefont{J.}~\bibnamefont{List}},
  \bibinfo{author}{\bibfnamefont{H.~E.} \bibnamefont{Logan}},
  \bibinfo{author}{\bibfnamefont{A.}~\bibnamefont{Nomerotski}},
  \bibinfo{author}{\bibfnamefont{M.}~\bibnamefont{Perelstein}},
  \bibnamefont{et~al.} (\bibinfo{year}{2013}), \eprint{1306.6352}.

\bibitem[{\citenamefont{Elias-Miro
  et~al.}(2013{\natexlab{a}})\citenamefont{Elias-Miro, Espinosa, Masso, and
  Pomarol}}]{Elias-Miro:2013gya}
\bibinfo{author}{\bibfnamefont{J.}~\bibnamefont{Elias-Miro}},
  \bibinfo{author}{\bibfnamefont{J.~R.} \bibnamefont{Espinosa}},
  \bibinfo{author}{\bibfnamefont{E.}~\bibnamefont{Masso}}, \bibnamefont{and}
  \bibinfo{author}{\bibfnamefont{A.}~\bibnamefont{Pomarol}},
  \bibinfo{journal}{JHEP} \textbf{\bibinfo{volume}{08}}, \bibinfo{pages}{033}
  (\bibinfo{year}{2013}{\natexlab{a}}), \eprint{1302.5661}.

\bibitem[{\citenamefont{Elias-Miro
  et~al.}(2013{\natexlab{b}})\citenamefont{Elias-Miro, Espinosa, Masso, and
  Pomarol}}]{Elias-Miro:2013mua}
\bibinfo{author}{\bibfnamefont{J.}~\bibnamefont{Elias-Miro}},
  \bibinfo{author}{\bibfnamefont{J.~R.} \bibnamefont{Espinosa}},
  \bibinfo{author}{\bibfnamefont{E.}~\bibnamefont{Masso}}, \bibnamefont{and}
  \bibinfo{author}{\bibfnamefont{A.}~\bibnamefont{Pomarol}},
  \bibinfo{journal}{JHEP} \textbf{\bibinfo{volume}{11}}, \bibinfo{pages}{066}
  (\bibinfo{year}{2013}{\natexlab{b}}), \eprint{1308.1879}.

\bibitem[{\citenamefont{Elias-Miro et~al.}(2014)\citenamefont{Elias-Miro,
  Grojean, Gupta, and Marzocca}}]{Elias-Miro:2013eta}
\bibinfo{author}{\bibfnamefont{J.}~\bibnamefont{Elias-Miro}},
  \bibinfo{author}{\bibfnamefont{C.}~\bibnamefont{Grojean}},
  \bibinfo{author}{\bibfnamefont{R.~S.} \bibnamefont{Gupta}}, \bibnamefont{and}
  \bibinfo{author}{\bibfnamefont{D.}~\bibnamefont{Marzocca}},
  \bibinfo{journal}{JHEP} \textbf{\bibinfo{volume}{05}}, \bibinfo{pages}{019}
  (\bibinfo{year}{2014}), \eprint{1312.2928}.

\bibitem[{\citenamefont{Jenkins et~al.}(2013)\citenamefont{Jenkins, Manohar,
  and Trott}}]{Jenkins:2013zja}
\bibinfo{author}{\bibfnamefont{E.~E.} \bibnamefont{Jenkins}},
  \bibinfo{author}{\bibfnamefont{A.~V.} \bibnamefont{Manohar}},
  \bibnamefont{and} \bibinfo{author}{\bibfnamefont{M.}~\bibnamefont{Trott}},
  \bibinfo{journal}{JHEP} \textbf{\bibinfo{volume}{10}}, \bibinfo{pages}{087}
  (\bibinfo{year}{2013}), \eprint{1308.2627}.

\bibitem[{\citenamefont{Jenkins et~al.}(2014)\citenamefont{Jenkins, Manohar,
  and Trott}}]{Jenkins:2013wua}
\bibinfo{author}{\bibfnamefont{E.~E.} \bibnamefont{Jenkins}},
  \bibinfo{author}{\bibfnamefont{A.~V.} \bibnamefont{Manohar}},
  \bibnamefont{and} \bibinfo{author}{\bibfnamefont{M.}~\bibnamefont{Trott}},
  \bibinfo{journal}{JHEP} \textbf{\bibinfo{volume}{01}}, \bibinfo{pages}{035}
  (\bibinfo{year}{2014}), \eprint{1310.4838}.

\bibitem[{\citenamefont{Alonso et~al.}(2014)\citenamefont{Alonso, Jenkins,
  Manohar, and Trott}}]{Alonso:2013hga}
\bibinfo{author}{\bibfnamefont{R.}~\bibnamefont{Alonso}},
  \bibinfo{author}{\bibfnamefont{E.~E.} \bibnamefont{Jenkins}},
  \bibinfo{author}{\bibfnamefont{A.~V.} \bibnamefont{Manohar}},
  \bibnamefont{and} \bibinfo{author}{\bibfnamefont{M.}~\bibnamefont{Trott}},
  \bibinfo{journal}{JHEP} \textbf{\bibinfo{volume}{04}}, \bibinfo{pages}{159}
  (\bibinfo{year}{2014}), \eprint{1312.2014}.

\bibitem[{\citenamefont{Passarino}(2013)}]{Passarino:2012cb}
\bibinfo{author}{\bibfnamefont{G.}~\bibnamefont{Passarino}},
  \bibinfo{journal}{Nucl. Phys.} \textbf{\bibinfo{volume}{B868}},
  \bibinfo{pages}{416} (\bibinfo{year}{2013}), \eprint{1209.5538}.

\bibitem[{\citenamefont{Chen et~al.}(2014)\citenamefont{Chen, Dawson, and
  Zhang}}]{Chen:2013kfa}
\bibinfo{author}{\bibfnamefont{C.-Y.} \bibnamefont{Chen}},
  \bibinfo{author}{\bibfnamefont{S.}~\bibnamefont{Dawson}}, \bibnamefont{and}
  \bibinfo{author}{\bibfnamefont{C.}~\bibnamefont{Zhang}},
  \bibinfo{journal}{Phys. Rev.} \textbf{\bibinfo{volume}{D89}},
  \bibinfo{pages}{015016} (\bibinfo{year}{2014}), \eprint{1311.3107}.

\bibitem[{\citenamefont{Grojean et~al.}(2013)\citenamefont{Grojean, Jenkins,
  Manohar, and Trott}}]{Grojean:2013kd}
\bibinfo{author}{\bibfnamefont{C.}~\bibnamefont{Grojean}},
  \bibinfo{author}{\bibfnamefont{E.~E.} \bibnamefont{Jenkins}},
  \bibinfo{author}{\bibfnamefont{A.~V.} \bibnamefont{Manohar}},
  \bibnamefont{and} \bibinfo{author}{\bibfnamefont{M.}~\bibnamefont{Trott}},
  \bibinfo{journal}{JHEP} \textbf{\bibinfo{volume}{04}}, \bibinfo{pages}{016}
  (\bibinfo{year}{2013}), \eprint{1301.2588}.

\bibitem[{\citenamefont{Englert and Spannowsky}(2015)}]{Englert:2014cva}
\bibinfo{author}{\bibfnamefont{C.}~\bibnamefont{Englert}} \bibnamefont{and}
  \bibinfo{author}{\bibfnamefont{M.}~\bibnamefont{Spannowsky}},
  \bibinfo{journal}{Phys. Lett.} \textbf{\bibinfo{volume}{B740}},
  \bibinfo{pages}{8} (\bibinfo{year}{2015}), \eprint{1408.5147}.

\bibitem[{\citenamefont{Zhang}(2014{\natexlab{a}})}]{Zhang:2014lya}
\bibinfo{author}{\bibfnamefont{C.}~\bibnamefont{Zhang}}, \bibinfo{journal}{J.
  Phys. Conf. Ser.} \textbf{\bibinfo{volume}{556}}, \bibinfo{pages}{012030}
  (\bibinfo{year}{2014}{\natexlab{a}}), \eprint{1410.2825}.

\bibitem[{\citenamefont{Pruna and Signer}(2014)}]{Pruna:2014asa}
\bibinfo{author}{\bibfnamefont{G.~M.} \bibnamefont{Pruna}} \bibnamefont{and}
  \bibinfo{author}{\bibfnamefont{A.}~\bibnamefont{Signer}},
  \bibinfo{journal}{JHEP} \textbf{\bibinfo{volume}{10}}, \bibinfo{pages}{14}
  (\bibinfo{year}{2014}), \eprint{1408.3565}.

\bibitem[{\citenamefont{Henning et~al.}(2014)\citenamefont{Henning, Lu, and
  Murayama}}]{Henning:2014wua}
\bibinfo{author}{\bibfnamefont{B.}~\bibnamefont{Henning}},
  \bibinfo{author}{\bibfnamefont{X.}~\bibnamefont{Lu}}, \bibnamefont{and}
  \bibinfo{author}{\bibfnamefont{H.}~\bibnamefont{Murayama}}
  (\bibinfo{year}{2014}), \eprint{1412.1837}.

\bibitem[{\citenamefont{Ghezzi et~al.}(2015)\citenamefont{Ghezzi,
  Gomez-Ambrosio, Passarino, and Uccirati}}]{Ghezzi:2015vva}
\bibinfo{author}{\bibfnamefont{M.}~\bibnamefont{Ghezzi}},
  \bibinfo{author}{\bibfnamefont{R.}~\bibnamefont{Gomez-Ambrosio}},
  \bibinfo{author}{\bibfnamefont{G.}~\bibnamefont{Passarino}},
  \bibnamefont{and} \bibinfo{author}{\bibfnamefont{S.}~\bibnamefont{Uccirati}},
  \bibinfo{journal}{JHEP} \textbf{\bibinfo{volume}{07}}, \bibinfo{pages}{175}
  (\bibinfo{year}{2015}), \eprint{1505.03706}.

\bibitem[{\citenamefont{David and Passarino}(2015)}]{David:2015waa}
\bibinfo{author}{\bibfnamefont{A.}~\bibnamefont{David}} \bibnamefont{and}
  \bibinfo{author}{\bibfnamefont{G.}~\bibnamefont{Passarino}}
  (\bibinfo{year}{2015}), \eprint{1510.00414}.

\bibitem[{\citenamefont{Grober et~al.}(2015)\citenamefont{Grober, Muhlleitner,
  Spira, and Streicher}}]{Grober:2015cwa}
\bibinfo{author}{\bibfnamefont{R.}~\bibnamefont{Grober}},
  \bibinfo{author}{\bibfnamefont{M.}~\bibnamefont{Muhlleitner}},
  \bibinfo{author}{\bibfnamefont{M.}~\bibnamefont{Spira}}, \bibnamefont{and}
  \bibinfo{author}{\bibfnamefont{J.}~\bibnamefont{Streicher}},
  \bibinfo{journal}{JHEP} \textbf{\bibinfo{volume}{09}}, \bibinfo{pages}{092}
  (\bibinfo{year}{2015}), \eprint{1504.06577}.

\bibitem[{\citenamefont{Hartmann and
  Trott}(2015{\natexlab{a}})}]{Hartmann:2015oia}
\bibinfo{author}{\bibfnamefont{C.}~\bibnamefont{Hartmann}} \bibnamefont{and}
  \bibinfo{author}{\bibfnamefont{M.}~\bibnamefont{Trott}},
  \bibinfo{journal}{JHEP} \textbf{\bibinfo{volume}{07}}, \bibinfo{pages}{151}
  (\bibinfo{year}{2015}{\natexlab{a}}), \eprint{1505.02646}.

\bibitem[{\citenamefont{Hartmann and
  Trott}(2015{\natexlab{b}})}]{Hartmann:2015aia}
\bibinfo{author}{\bibfnamefont{C.}~\bibnamefont{Hartmann}} \bibnamefont{and}
  \bibinfo{author}{\bibfnamefont{M.}~\bibnamefont{Trott}},
  \bibinfo{journal}{Phys. Rev. Lett.} \textbf{\bibinfo{volume}{115}},
  \bibinfo{pages}{191801} (\bibinfo{year}{2015}{\natexlab{b}}),
  \eprint{1507.03568}.

\bibitem[{\citenamefont{Gauld et~al.}(2016)\citenamefont{Gauld, Pecjak, and
  Scott}}]{Gauld:2015lmb}
\bibinfo{author}{\bibfnamefont{R.}~\bibnamefont{Gauld}},
  \bibinfo{author}{\bibfnamefont{B.~D.} \bibnamefont{Pecjak}},
  \bibnamefont{and} \bibinfo{author}{\bibfnamefont{D.~J.} \bibnamefont{Scott}},
  \bibinfo{journal}{JHEP} \textbf{\bibinfo{volume}{05}}, \bibinfo{pages}{080}
  (\bibinfo{year}{2016}), \eprint{1512.02508}.

\bibitem[{\citenamefont{Zhang}(2016)}]{Zhang:2016omx}
\bibinfo{author}{\bibfnamefont{C.}~\bibnamefont{Zhang}},
  \bibinfo{journal}{Phys. Rev. Lett.} \textbf{\bibinfo{volume}{116}},
  \bibinfo{pages}{162002} (\bibinfo{year}{2016}), \eprint{1601.06163}.

\bibitem[{\citenamefont{Maltoni et~al.}(2016)\citenamefont{Maltoni, Vryonidou,
  and Zhang}}]{Maltoni:2016yxb}
\bibinfo{author}{\bibfnamefont{F.}~\bibnamefont{Maltoni}},
  \bibinfo{author}{\bibfnamefont{E.}~\bibnamefont{Vryonidou}},
  \bibnamefont{and} \bibinfo{author}{\bibfnamefont{C.}~\bibnamefont{Zhang}}
  (\bibinfo{year}{2016}), \eprint{1607.05330}.

\bibitem[{\citenamefont{Zhang and Maltoni}(2013)}]{Zhang:2013xya}
\bibinfo{author}{\bibfnamefont{C.}~\bibnamefont{Zhang}} \bibnamefont{and}
  \bibinfo{author}{\bibfnamefont{F.}~\bibnamefont{Maltoni}},
  \bibinfo{journal}{Phys. Rev.} \textbf{\bibinfo{volume}{D88}},
  \bibinfo{pages}{054005} (\bibinfo{year}{2013}), \eprint{1305.7386}.

\bibitem[{\citenamefont{Zhang}(2014{\natexlab{b}})}]{Zhang:2014rja}
\bibinfo{author}{\bibfnamefont{C.}~\bibnamefont{Zhang}},
  \bibinfo{journal}{Phys. Rev.} \textbf{\bibinfo{volume}{D90}},
  \bibinfo{pages}{014008} (\bibinfo{year}{2014}{\natexlab{b}}),
  \eprint{1404.1264}.

\bibitem[{\citenamefont{Degrande et~al.}(2015)\citenamefont{Degrande, Maltoni,
  Wang, and Zhang}}]{Degrande:2014tta}
\bibinfo{author}{\bibfnamefont{C.}~\bibnamefont{Degrande}},
  \bibinfo{author}{\bibfnamefont{F.}~\bibnamefont{Maltoni}},
  \bibinfo{author}{\bibfnamefont{J.}~\bibnamefont{Wang}}, \bibnamefont{and}
  \bibinfo{author}{\bibfnamefont{C.}~\bibnamefont{Zhang}},
  \bibinfo{journal}{Phys. Rev.} \textbf{\bibinfo{volume}{D91}},
  \bibinfo{pages}{034024} (\bibinfo{year}{2015}), \eprint{1412.5594}.

\bibitem[{\citenamefont{Buarque~Franzosi and Zhang}(2015)}]{Franzosi:2015osa}
\bibinfo{author}{\bibfnamefont{D.}~\bibnamefont{Buarque~Franzosi}}
  \bibnamefont{and} \bibinfo{author}{\bibfnamefont{C.}~\bibnamefont{Zhang}},
  \bibinfo{journal}{Phys. Rev.} \textbf{\bibinfo{volume}{D91}},
  \bibinfo{pages}{114010} (\bibinfo{year}{2015}), \eprint{1503.08841}.

\bibitem[{\citenamefont{Bessidskaia~Bylund
  et~al.}(2016)\citenamefont{Bessidskaia~Bylund, Maltoni, Tsinikos, Vryonidou,
  and Zhang}}]{Bylund:2016phk}
\bibinfo{author}{\bibfnamefont{O.}~\bibnamefont{Bessidskaia~Bylund}},
  \bibinfo{author}{\bibfnamefont{F.}~\bibnamefont{Maltoni}},
  \bibinfo{author}{\bibfnamefont{I.}~\bibnamefont{Tsinikos}},
  \bibinfo{author}{\bibfnamefont{E.}~\bibnamefont{Vryonidou}},
  \bibnamefont{and} \bibinfo{author}{\bibfnamefont{C.}~\bibnamefont{Zhang}},
  \bibinfo{journal}{JHEP} \textbf{\bibinfo{volume}{05}}, \bibinfo{pages}{052}
  (\bibinfo{year}{2016}), \eprint{1601.08193}.

\bibitem[{\citenamefont{Gorbahn and Haisch}(2016)}]{Gorbahn:2016uoy}
\bibinfo{author}{\bibfnamefont{M.}~\bibnamefont{Gorbahn}} \bibnamefont{and}
  \bibinfo{author}{\bibfnamefont{U.}~\bibnamefont{Haisch}}
  (\bibinfo{year}{2016}), \eprint{1607.03773}.

\bibitem[{\citenamefont{Grzadkowski et~al.}(2010)\citenamefont{Grzadkowski,
  Iskrzynski, Misiak, and Rosiek}}]{Grzadkowski:2010es}
\bibinfo{author}{\bibfnamefont{B.}~\bibnamefont{Grzadkowski}},
  \bibinfo{author}{\bibfnamefont{M.}~\bibnamefont{Iskrzynski}},
  \bibinfo{author}{\bibfnamefont{M.}~\bibnamefont{Misiak}}, \bibnamefont{and}
  \bibinfo{author}{\bibfnamefont{J.}~\bibnamefont{Rosiek}},
  \bibinfo{journal}{JHEP} \textbf{\bibinfo{volume}{10}}, \bibinfo{pages}{085}
  (\bibinfo{year}{2010}), \eprint{1008.4884}.

\bibitem[{\citenamefont{Wells and Zhang}(2015)}]{Wells:2015cre}
\bibinfo{author}{\bibfnamefont{J.~D.} \bibnamefont{Wells}} \bibnamefont{and}
  \bibinfo{author}{\bibfnamefont{Z.}~\bibnamefont{Zhang}}
  (\bibinfo{year}{2015}), \eprint{1512.03056}.

\bibitem[{\citenamefont{Chivukula and Georgi}(1987)}]{Chivukula:1987py}
\bibinfo{author}{\bibfnamefont{R.~S.} \bibnamefont{Chivukula}}
  \bibnamefont{and} \bibinfo{author}{\bibfnamefont{H.}~\bibnamefont{Georgi}},
  \bibinfo{journal}{Phys. Lett.} \textbf{\bibinfo{volume}{B188}},
  \bibinfo{pages}{99} (\bibinfo{year}{1987}).

\bibitem[{\citenamefont{D'Ambrosio et~al.}(2002)\citenamefont{D'Ambrosio,
  Giudice, Isidori, and Strumia}}]{D'Ambrosio:2002ex}
\bibinfo{author}{\bibfnamefont{G.}~\bibnamefont{D'Ambrosio}},
  \bibinfo{author}{\bibfnamefont{G.~F.} \bibnamefont{Giudice}},
  \bibinfo{author}{\bibfnamefont{G.}~\bibnamefont{Isidori}}, \bibnamefont{and}
  \bibinfo{author}{\bibfnamefont{A.}~\bibnamefont{Strumia}},
  \bibinfo{journal}{Nucl. Phys.} \textbf{\bibinfo{volume}{B645}},
  \bibinfo{pages}{155} (\bibinfo{year}{2002}), \eprint{hep-ph/0207036}.

\bibitem[{\citenamefont{Denner}(1993)}]{Denner:1991kt}
\bibinfo{author}{\bibfnamefont{A.}~\bibnamefont{Denner}},
  \bibinfo{journal}{Fortsch. Phys.} \textbf{\bibinfo{volume}{41}},
  \bibinfo{pages}{307} (\bibinfo{year}{1993}), \eprint{0709.1075}.

\bibitem[{\citenamefont{Kniehl and Pilaftsis}(1996)}]{Kniehl:1996bd}
\bibinfo{author}{\bibfnamefont{B.~A.} \bibnamefont{Kniehl}} \bibnamefont{and}
  \bibinfo{author}{\bibfnamefont{A.}~\bibnamefont{Pilaftsis}},
  \bibinfo{journal}{Nucl. Phys.} \textbf{\bibinfo{volume}{B474}},
  \bibinfo{pages}{286} (\bibinfo{year}{1996}), \eprint{hep-ph/9601390}.

\bibitem[{\citenamefont{Alloul et~al.}(2014)\citenamefont{Alloul, Christensen,
  Degrande, Duhr, and Fuks}}]{Alloul:2013bka}
\bibinfo{author}{\bibfnamefont{A.}~\bibnamefont{Alloul}},
  \bibinfo{author}{\bibfnamefont{N.~D.} \bibnamefont{Christensen}},
  \bibinfo{author}{\bibfnamefont{C.}~\bibnamefont{Degrande}},
  \bibinfo{author}{\bibfnamefont{C.}~\bibnamefont{Duhr}}, \bibnamefont{and}
  \bibinfo{author}{\bibfnamefont{B.}~\bibnamefont{Fuks}},
  \bibinfo{journal}{Comput. Phys. Commun.} \textbf{\bibinfo{volume}{185}},
  \bibinfo{pages}{2250} (\bibinfo{year}{2014}), \eprint{1310.1921}.

\bibitem[{\citenamefont{Hahn}(2001)}]{Hahn:2000kx}
\bibinfo{author}{\bibfnamefont{T.}~\bibnamefont{Hahn}},
  \bibinfo{journal}{Comput. Phys. Commun.} \textbf{\bibinfo{volume}{140}},
  \bibinfo{pages}{418} (\bibinfo{year}{2001}), \eprint{hep-ph/0012260}.

\bibitem[{\citenamefont{Hahn and Perez-Victoria}(1999)}]{Hahn:1998yk}
\bibinfo{author}{\bibfnamefont{T.}~\bibnamefont{Hahn}} \bibnamefont{and}
  \bibinfo{author}{\bibfnamefont{M.}~\bibnamefont{Perez-Victoria}},
  \bibinfo{journal}{Comput. Phys. Commun.} \textbf{\bibinfo{volume}{118}},
  \bibinfo{pages}{153} (\bibinfo{year}{1999}), \eprint{hep-ph/9807565}.

\bibitem[{\citenamefont{Braaten and Leveille}(1980)}]{Braaten:1980yq}
\bibinfo{author}{\bibfnamefont{E.}~\bibnamefont{Braaten}} \bibnamefont{and}
  \bibinfo{author}{\bibfnamefont{J.~P.} \bibnamefont{Leveille}},
  \bibinfo{journal}{Phys. Rev.} \textbf{\bibinfo{volume}{D22}},
  \bibinfo{pages}{715} (\bibinfo{year}{1980}).

\bibitem[{\citenamefont{Sakai}(1980)}]{Sakai:1980fa}
\bibinfo{author}{\bibfnamefont{N.}~\bibnamefont{Sakai}},
  \bibinfo{journal}{Phys. Rev.} \textbf{\bibinfo{volume}{D22}},
  \bibinfo{pages}{2220} (\bibinfo{year}{1980}).

\bibitem[{\citenamefont{Inami and Kubota}(1981)}]{Inami:1980qp}
\bibinfo{author}{\bibfnamefont{T.}~\bibnamefont{Inami}} \bibnamefont{and}
  \bibinfo{author}{\bibfnamefont{T.}~\bibnamefont{Kubota}},
  \bibinfo{journal}{Nucl. Phys.} \textbf{\bibinfo{volume}{B179}},
  \bibinfo{pages}{171} (\bibinfo{year}{1981}).

\bibitem[{\citenamefont{Drees and Hikasa}(1990{\natexlab{a}})}]{Drees:1989du}
\bibinfo{author}{\bibfnamefont{M.}~\bibnamefont{Drees}} \bibnamefont{and}
  \bibinfo{author}{\bibfnamefont{K.-i.} \bibnamefont{Hikasa}},
  \bibinfo{journal}{Phys. Rev.} \textbf{\bibinfo{volume}{D41}},
  \bibinfo{pages}{1547} (\bibinfo{year}{1990}{\natexlab{a}}).

\bibitem[{\citenamefont{Drees and Hikasa}(1990{\natexlab{b}})}]{Drees:1990dq}
\bibinfo{author}{\bibfnamefont{M.}~\bibnamefont{Drees}} \bibnamefont{and}
  \bibinfo{author}{\bibfnamefont{K.-i.} \bibnamefont{Hikasa}},
  \bibinfo{journal}{Phys. Lett.} \textbf{\bibinfo{volume}{B240}},
  \bibinfo{pages}{455} (\bibinfo{year}{1990}{\natexlab{b}}),
  \bibinfo{note}{[Erratum: Phys. Lett.B262,497(1991)]}.

\bibitem[{\citenamefont{Azatov et~al.}(2012)\citenamefont{Azatov, Contino, and
  Galloway}}]{Azatov:2012bz}
\bibinfo{author}{\bibfnamefont{A.}~\bibnamefont{Azatov}},
  \bibinfo{author}{\bibfnamefont{R.}~\bibnamefont{Contino}}, \bibnamefont{and}
  \bibinfo{author}{\bibfnamefont{J.}~\bibnamefont{Galloway}},
  \bibinfo{journal}{JHEP} \textbf{\bibinfo{volume}{04}}, \bibinfo{pages}{127}
  (\bibinfo{year}{2012}), \bibinfo{note}{[Erratum: JHEP04,140(2013)]},
  \eprint{1202.3415}.

\bibitem[{\citenamefont{Espinosa et~al.}(2012)\citenamefont{Espinosa, Grojean,
  Muhlleitner, and Trott}}]{Espinosa:2012im}
\bibinfo{author}{\bibfnamefont{J.~R.} \bibnamefont{Espinosa}},
  \bibinfo{author}{\bibfnamefont{C.}~\bibnamefont{Grojean}},
  \bibinfo{author}{\bibfnamefont{M.}~\bibnamefont{Muhlleitner}},
  \bibnamefont{and} \bibinfo{author}{\bibfnamefont{M.}~\bibnamefont{Trott}},
  \bibinfo{journal}{JHEP} \textbf{\bibinfo{volume}{12}}, \bibinfo{pages}{045}
  (\bibinfo{year}{2012}), \eprint{1207.1717}.

\bibitem[{\citenamefont{Plehn and Rauch}(2012)}]{Plehn:2012iz}
\bibinfo{author}{\bibfnamefont{T.}~\bibnamefont{Plehn}} \bibnamefont{and}
  \bibinfo{author}{\bibfnamefont{M.}~\bibnamefont{Rauch}},
  \bibinfo{journal}{Europhys. Lett.} \textbf{\bibinfo{volume}{100}},
  \bibinfo{pages}{11002} (\bibinfo{year}{2012}), \eprint{1207.6108}.

\bibitem[{\citenamefont{Carmi et~al.}(2012)\citenamefont{Carmi, Falkowski,
  Kuflik, Volansky, and Zupan}}]{Carmi:2012in}
\bibinfo{author}{\bibfnamefont{D.}~\bibnamefont{Carmi}},
  \bibinfo{author}{\bibfnamefont{A.}~\bibnamefont{Falkowski}},
  \bibinfo{author}{\bibfnamefont{E.}~\bibnamefont{Kuflik}},
  \bibinfo{author}{\bibfnamefont{T.}~\bibnamefont{Volansky}}, \bibnamefont{and}
  \bibinfo{author}{\bibfnamefont{J.}~\bibnamefont{Zupan}},
  \bibinfo{journal}{JHEP} \textbf{\bibinfo{volume}{10}}, \bibinfo{pages}{196}
  (\bibinfo{year}{2012}), \eprint{1207.1718}.

\bibitem[{\citenamefont{Peskin}(2012)}]{Peskin:2012we}
\bibinfo{author}{\bibfnamefont{M.~E.} \bibnamefont{Peskin}}
  (\bibinfo{year}{2012}), \eprint{1207.2516}.

\bibitem[{\citenamefont{Corbett et~al.}(2013)\citenamefont{Corbett, Eboli,
  Gonzalez-Fraile, and Gonzalez-Garcia}}]{Corbett:2012ja}
\bibinfo{author}{\bibfnamefont{T.}~\bibnamefont{Corbett}},
  \bibinfo{author}{\bibfnamefont{O.~J.~P.} \bibnamefont{Eboli}},
  \bibinfo{author}{\bibfnamefont{J.}~\bibnamefont{Gonzalez-Fraile}},
  \bibnamefont{and} \bibinfo{author}{\bibfnamefont{M.~C.}
  \bibnamefont{Gonzalez-Garcia}}, \bibinfo{journal}{Phys. Rev.}
  \textbf{\bibinfo{volume}{D87}}, \bibinfo{pages}{015022}
  (\bibinfo{year}{2013}), \eprint{1211.4580}.

\bibitem[{\citenamefont{Masso and Sanz}(2013)}]{Masso:2012eq}
\bibinfo{author}{\bibfnamefont{E.}~\bibnamefont{Masso}} \bibnamefont{and}
  \bibinfo{author}{\bibfnamefont{V.}~\bibnamefont{Sanz}},
  \bibinfo{journal}{Phys. Rev.} \textbf{\bibinfo{volume}{D87}},
  \bibinfo{pages}{033001} (\bibinfo{year}{2013}), \eprint{1211.1320}.

\bibitem[{\citenamefont{Falkowski et~al.}(2013)\citenamefont{Falkowski, Riva,
  and Urbano}}]{Falkowski:2013dza}
\bibinfo{author}{\bibfnamefont{A.}~\bibnamefont{Falkowski}},
  \bibinfo{author}{\bibfnamefont{F.}~\bibnamefont{Riva}}, \bibnamefont{and}
  \bibinfo{author}{\bibfnamefont{A.}~\bibnamefont{Urbano}},
  \bibinfo{journal}{JHEP} \textbf{\bibinfo{volume}{11}}, \bibinfo{pages}{111}
  (\bibinfo{year}{2013}), \eprint{1303.1812}.

\bibitem[{\citenamefont{Dumont et~al.}(2013)\citenamefont{Dumont, Fichet, and
  von Gersdorff}}]{Dumont:2013wma}
\bibinfo{author}{\bibfnamefont{B.}~\bibnamefont{Dumont}},
  \bibinfo{author}{\bibfnamefont{S.}~\bibnamefont{Fichet}}, \bibnamefont{and}
  \bibinfo{author}{\bibfnamefont{G.}~\bibnamefont{von Gersdorff}},
  \bibinfo{journal}{JHEP} \textbf{\bibinfo{volume}{07}}, \bibinfo{pages}{065}
  (\bibinfo{year}{2013}), \eprint{1304.3369}.

\bibitem[{\citenamefont{Djouadi and Moreau}(2013)}]{Djouadi:2013qya}
\bibinfo{author}{\bibfnamefont{A.}~\bibnamefont{Djouadi}} \bibnamefont{and}
  \bibinfo{author}{\bibfnamefont{G.}~\bibnamefont{Moreau}},
  \bibinfo{journal}{Eur. Phys. J.} \textbf{\bibinfo{volume}{C73}},
  \bibinfo{pages}{2512} (\bibinfo{year}{2013}), \eprint{1303.6591}.

\bibitem[{\citenamefont{Lopez-Val et~al.}(2013)\citenamefont{Lopez-Val, Plehn,
  and Rauch}}]{Lopez-Val:2013yba}
\bibinfo{author}{\bibfnamefont{D.}~\bibnamefont{Lopez-Val}},
  \bibinfo{author}{\bibfnamefont{T.}~\bibnamefont{Plehn}}, \bibnamefont{and}
  \bibinfo{author}{\bibfnamefont{M.}~\bibnamefont{Rauch}},
  \bibinfo{journal}{JHEP} \textbf{\bibinfo{volume}{10}}, \bibinfo{pages}{134}
  (\bibinfo{year}{2013}), \eprint{1308.1979}.

\bibitem[{\citenamefont{Pomarol and Riva}(2014)}]{Pomarol:2013zra}
\bibinfo{author}{\bibfnamefont{A.}~\bibnamefont{Pomarol}} \bibnamefont{and}
  \bibinfo{author}{\bibfnamefont{F.}~\bibnamefont{Riva}},
  \bibinfo{journal}{JHEP} \textbf{\bibinfo{volume}{01}}, \bibinfo{pages}{151}
  (\bibinfo{year}{2014}), \eprint{1308.2803}.

\bibitem[{\citenamefont{Englert et~al.}(2014)\citenamefont{Englert, Freitas,
  Muhlleitner, Plehn, Rauch, Spira, and Walz}}]{Englert:2014uua}
\bibinfo{author}{\bibfnamefont{C.}~\bibnamefont{Englert}},
  \bibinfo{author}{\bibfnamefont{A.}~\bibnamefont{Freitas}},
  \bibinfo{author}{\bibfnamefont{M.~M.} \bibnamefont{Muhlleitner}},
  \bibinfo{author}{\bibfnamefont{T.}~\bibnamefont{Plehn}},
  \bibinfo{author}{\bibfnamefont{M.}~\bibnamefont{Rauch}},
  \bibinfo{author}{\bibfnamefont{M.}~\bibnamefont{Spira}}, \bibnamefont{and}
  \bibinfo{author}{\bibfnamefont{K.}~\bibnamefont{Walz}}, \bibinfo{journal}{J.
  Phys.} \textbf{\bibinfo{volume}{G41}}, \bibinfo{pages}{113001}
  (\bibinfo{year}{2014}), \eprint{1403.7191}.

\bibitem[{\citenamefont{Ellis et~al.}(2014)\citenamefont{Ellis, Sanz, and
  You}}]{Ellis:2014dva}
\bibinfo{author}{\bibfnamefont{J.}~\bibnamefont{Ellis}},
  \bibinfo{author}{\bibfnamefont{V.}~\bibnamefont{Sanz}}, \bibnamefont{and}
  \bibinfo{author}{\bibfnamefont{T.}~\bibnamefont{You}},
  \bibinfo{journal}{JHEP} \textbf{\bibinfo{volume}{07}}, \bibinfo{pages}{036}
  (\bibinfo{year}{2014}), \eprint{1404.3667}.

\bibitem[{\citenamefont{Belusca-Maito}(2014)}]{Belusca-Maito:2014dpa}
\bibinfo{author}{\bibfnamefont{H.}~\bibnamefont{Belusca-Maito}}
  (\bibinfo{year}{2014}), \eprint{1404.5343}.

\bibitem[{\citenamefont{Ellis et~al.}(2015)\citenamefont{Ellis, Sanz, and
  You}}]{Ellis:2014jta}
\bibinfo{author}{\bibfnamefont{J.}~\bibnamefont{Ellis}},
  \bibinfo{author}{\bibfnamefont{V.}~\bibnamefont{Sanz}}, \bibnamefont{and}
  \bibinfo{author}{\bibfnamefont{T.}~\bibnamefont{You}},
  \bibinfo{journal}{JHEP} \textbf{\bibinfo{volume}{03}}, \bibinfo{pages}{157}
  (\bibinfo{year}{2015}), \eprint{1410.7703}.

\bibitem[{\citenamefont{Corbett et~al.}(2015)\citenamefont{Corbett, Eboli,
  Goncalves, Gonzalez-Fraile, Plehn, and Rauch}}]{Corbett:2015ksa}
\bibinfo{author}{\bibfnamefont{T.}~\bibnamefont{Corbett}},
  \bibinfo{author}{\bibfnamefont{O.~J.~P.} \bibnamefont{Eboli}},
  \bibinfo{author}{\bibfnamefont{D.}~\bibnamefont{Goncalves}},
  \bibinfo{author}{\bibfnamefont{J.}~\bibnamefont{Gonzalez-Fraile}},
  \bibinfo{author}{\bibfnamefont{T.}~\bibnamefont{Plehn}}, \bibnamefont{and}
  \bibinfo{author}{\bibfnamefont{M.}~\bibnamefont{Rauch}},
  \bibinfo{journal}{JHEP} \textbf{\bibinfo{volume}{08}}, \bibinfo{pages}{156}
  (\bibinfo{year}{2015}), \eprint{1505.05516}.

\bibitem[{\citenamefont{Buchalla et~al.}(2016)\citenamefont{Buchalla, Cata,
  Celis, and Krause}}]{Buchalla:2015qju}
\bibinfo{author}{\bibfnamefont{G.}~\bibnamefont{Buchalla}},
  \bibinfo{author}{\bibfnamefont{O.}~\bibnamefont{Cata}},
  \bibinfo{author}{\bibfnamefont{A.}~\bibnamefont{Celis}}, \bibnamefont{and}
  \bibinfo{author}{\bibfnamefont{C.}~\bibnamefont{Krause}},
  \bibinfo{journal}{Eur. Phys. J.} \textbf{\bibinfo{volume}{C76}},
  \bibinfo{pages}{233} (\bibinfo{year}{2016}), \eprint{1511.00988}.

\bibitem[{\citenamefont{Englert et~al.}(2015)\citenamefont{Englert, Kogler,
  Schulz, and Spannowsky}}]{Englert:2015hrx}
\bibinfo{author}{\bibfnamefont{C.}~\bibnamefont{Englert}},
  \bibinfo{author}{\bibfnamefont{R.}~\bibnamefont{Kogler}},
  \bibinfo{author}{\bibfnamefont{H.}~\bibnamefont{Schulz}}, \bibnamefont{and}
  \bibinfo{author}{\bibfnamefont{M.}~\bibnamefont{Spannowsky}}
  (\bibinfo{year}{2015}), \eprint{1511.05170}.

\bibitem[{\citenamefont{Aad et~al.}(2016)}]{Aad:2015tna}
\bibinfo{author}{\bibfnamefont{G.}~\bibnamefont{Aad}} \bibnamefont{et~al.}
  (\bibinfo{collaboration}{ATLAS}), \bibinfo{journal}{Phys. Lett.}
  \textbf{\bibinfo{volume}{B753}}, \bibinfo{pages}{69} (\bibinfo{year}{2016}),
  \eprint{1508.02507}.

\bibitem[{\citenamefont{Butter et~al.}(2016)\citenamefont{Butter, Eboli,
  Gonzalez-Fraile, Gonzalez-Garcia, Plehn, and Rauch}}]{Butter:2016cvz}
\bibinfo{author}{\bibfnamefont{A.}~\bibnamefont{Butter}},
  \bibinfo{author}{\bibfnamefont{O.~J.~P.} \bibnamefont{Eboli}},
  \bibinfo{author}{\bibfnamefont{J.}~\bibnamefont{Gonzalez-Fraile}},
  \bibinfo{author}{\bibfnamefont{M.~C.} \bibnamefont{Gonzalez-Garcia}},
  \bibinfo{author}{\bibfnamefont{T.}~\bibnamefont{Plehn}}, \bibnamefont{and}
  \bibinfo{author}{\bibfnamefont{M.}~\bibnamefont{Rauch}}
  (\bibinfo{year}{2016}), \eprint{1604.03105}.

\bibitem[{\citenamefont{Cirigliano et~al.}(2016)\citenamefont{Cirigliano,
  Dekens, de~Vries, and Mereghetti}}]{Cirigliano:2016nyn}
\bibinfo{author}{\bibfnamefont{V.}~\bibnamefont{Cirigliano}},
  \bibinfo{author}{\bibfnamefont{W.}~\bibnamefont{Dekens}},
  \bibinfo{author}{\bibfnamefont{J.}~\bibnamefont{de~Vries}}, \bibnamefont{and}
  \bibinfo{author}{\bibfnamefont{E.}~\bibnamefont{Mereghetti}}
  (\bibinfo{year}{2016}), \eprint{1605.04311}.

\bibitem[{\citenamefont{Contino et~al.}(2013)\citenamefont{Contino, Ghezzi,
  Grojean, Muhlleitner, and Spira}}]{Contino:2013kra}
\bibinfo{author}{\bibfnamefont{R.}~\bibnamefont{Contino}},
  \bibinfo{author}{\bibfnamefont{M.}~\bibnamefont{Ghezzi}},
  \bibinfo{author}{\bibfnamefont{C.}~\bibnamefont{Grojean}},
  \bibinfo{author}{\bibfnamefont{M.}~\bibnamefont{Muhlleitner}},
  \bibnamefont{and} \bibinfo{author}{\bibfnamefont{M.}~\bibnamefont{Spira}},
  \bibinfo{journal}{JHEP} \textbf{\bibinfo{volume}{07}}, \bibinfo{pages}{035}
  (\bibinfo{year}{2013}), \eprint{1303.3876}.

\bibitem[{\citenamefont{Contino et~al.}(2014)\citenamefont{Contino, Ghezzi,
  Grojean, Muhlleitner, and Spira}}]{Contino:2014aaa}
\bibinfo{author}{\bibfnamefont{R.}~\bibnamefont{Contino}},
  \bibinfo{author}{\bibfnamefont{M.}~\bibnamefont{Ghezzi}},
  \bibinfo{author}{\bibfnamefont{C.}~\bibnamefont{Grojean}},
  \bibinfo{author}{\bibfnamefont{M.}~\bibnamefont{Muhlleitner}},
  \bibnamefont{and} \bibinfo{author}{\bibfnamefont{M.}~\bibnamefont{Spira}},
  \bibinfo{journal}{Comput. Phys. Commun.} \textbf{\bibinfo{volume}{185}},
  \bibinfo{pages}{3412} (\bibinfo{year}{2014}), \eprint{1403.3381}.

\bibitem[{\citenamefont{Giudice et~al.}(2007)\citenamefont{Giudice, Grojean,
  Pomarol, and Rattazzi}}]{Giudice:2007fh}
\bibinfo{author}{\bibfnamefont{G.~F.} \bibnamefont{Giudice}},
  \bibinfo{author}{\bibfnamefont{C.}~\bibnamefont{Grojean}},
  \bibinfo{author}{\bibfnamefont{A.}~\bibnamefont{Pomarol}}, \bibnamefont{and}
  \bibinfo{author}{\bibfnamefont{R.}~\bibnamefont{Rattazzi}},
  \bibinfo{journal}{JHEP} \textbf{\bibinfo{volume}{06}}, \bibinfo{pages}{045}
  (\bibinfo{year}{2007}), \eprint{hep-ph/0703164}.

\bibitem[{\citenamefont{Djouadi et~al.}(1998)\citenamefont{Djouadi, Kalinowski,
  and Spira}}]{Djouadi:1997yw}
\bibinfo{author}{\bibfnamefont{A.}~\bibnamefont{Djouadi}},
  \bibinfo{author}{\bibfnamefont{J.}~\bibnamefont{Kalinowski}},
  \bibnamefont{and} \bibinfo{author}{\bibfnamefont{M.}~\bibnamefont{Spira}},
  \bibinfo{journal}{Comput. Phys. Commun.} \textbf{\bibinfo{volume}{108}},
  \bibinfo{pages}{56} (\bibinfo{year}{1998}), \eprint{hep-ph/9704448}.

\bibitem[{\citenamefont{Butterworth et~al.}(2010)}]{Butterworth:2010ym}
\bibinfo{author}{\bibfnamefont{J.~M.} \bibnamefont{Butterworth}}
  \bibnamefont{et~al.} (\bibinfo{year}{2010}), \eprint{1003.1643},
  \urlprefix\url{https://inspirehep.net/record/848006/files/arXiv:1003.1643.pdf}.

\bibitem[{\citenamefont{Gorishnii et~al.}(1984)\citenamefont{Gorishnii, Kataev,
  and Larin}}]{Gorishnii:1983cu}
\bibinfo{author}{\bibfnamefont{S.~G.} \bibnamefont{Gorishnii}},
  \bibinfo{author}{\bibfnamefont{A.~L.} \bibnamefont{Kataev}},
  \bibnamefont{and} \bibinfo{author}{\bibfnamefont{S.~A.} \bibnamefont{Larin}},
  \bibinfo{journal}{Sov. J. Nucl. Phys.} \textbf{\bibinfo{volume}{40}},
  \bibinfo{pages}{329} (\bibinfo{year}{1984}), \bibinfo{note}{[Yad.
  Fiz.40,517(1984)]}.

\bibitem[{\citenamefont{Gorishnii et~al.}(1991)\citenamefont{Gorishnii, Kataev,
  Larin, and Surguladze}}]{Gorishnii:1991zr}
\bibinfo{author}{\bibfnamefont{S.~G.} \bibnamefont{Gorishnii}},
  \bibinfo{author}{\bibfnamefont{A.~L.} \bibnamefont{Kataev}},
  \bibinfo{author}{\bibfnamefont{S.~A.} \bibnamefont{Larin}}, \bibnamefont{and}
  \bibinfo{author}{\bibfnamefont{L.~R.} \bibnamefont{Surguladze}},
  \bibinfo{journal}{Phys. Rev.} \textbf{\bibinfo{volume}{D43}},
  \bibinfo{pages}{1633} (\bibinfo{year}{1991}).

\bibitem[{\citenamefont{Kataev and Kim}(1994)}]{Kataev:1993be}
\bibinfo{author}{\bibfnamefont{A.~L.} \bibnamefont{Kataev}} \bibnamefont{and}
  \bibinfo{author}{\bibfnamefont{V.~T.} \bibnamefont{Kim}},
  \bibinfo{journal}{Mod. Phys. Lett.} \textbf{\bibinfo{volume}{A9}},
  \bibinfo{pages}{1309} (\bibinfo{year}{1994}).

\bibitem[{\citenamefont{Surguladze}(1994)}]{Surguladze:1994gc}
\bibinfo{author}{\bibfnamefont{L.~R.} \bibnamefont{Surguladze}},
  \bibinfo{journal}{Phys. Lett.} \textbf{\bibinfo{volume}{B341}},
  \bibinfo{pages}{60} (\bibinfo{year}{1994}), \eprint{hep-ph/9405325}.

\bibitem[{\citenamefont{Larin et~al.}(1995)\citenamefont{Larin, van Ritbergen,
  and Vermaseren}}]{Larin:1995sq}
\bibinfo{author}{\bibfnamefont{S.~A.} \bibnamefont{Larin}},
  \bibinfo{author}{\bibfnamefont{T.}~\bibnamefont{van Ritbergen}},
  \bibnamefont{and} \bibinfo{author}{\bibfnamefont{J.~A.~M.}
  \bibnamefont{Vermaseren}}, \bibinfo{journal}{Phys. Lett.}
  \textbf{\bibinfo{volume}{B362}}, \bibinfo{pages}{134} (\bibinfo{year}{1995}),
  \eprint{hep-ph/9506465}.

\bibitem[{\citenamefont{Chetyrkin and Kwiatkowski}(1996)}]{Chetyrkin:1995pd}
\bibinfo{author}{\bibfnamefont{K.~G.} \bibnamefont{Chetyrkin}}
  \bibnamefont{and}
  \bibinfo{author}{\bibfnamefont{A.}~\bibnamefont{Kwiatkowski}},
  \bibinfo{journal}{Nucl. Phys.} \textbf{\bibinfo{volume}{B461}},
  \bibinfo{pages}{3} (\bibinfo{year}{1996}), \eprint{hep-ph/9505358}.

\bibitem[{\citenamefont{Chetyrkin}(1997)}]{Chetyrkin:1996sr}
\bibinfo{author}{\bibfnamefont{K.~G.} \bibnamefont{Chetyrkin}},
  \bibinfo{journal}{Phys. Lett.} \textbf{\bibinfo{volume}{B390}},
  \bibinfo{pages}{309} (\bibinfo{year}{1997}), \eprint{hep-ph/9608318}.

\bibitem[{\citenamefont{Baikov et~al.}(2006)\citenamefont{Baikov, Chetyrkin,
  and Kuhn}}]{Baikov:2005rw}
\bibinfo{author}{\bibfnamefont{P.~A.} \bibnamefont{Baikov}},
  \bibinfo{author}{\bibfnamefont{K.~G.} \bibnamefont{Chetyrkin}},
  \bibnamefont{and} \bibinfo{author}{\bibfnamefont{J.~H.} \bibnamefont{Kuhn}},
  \bibinfo{journal}{Phys. Rev. Lett.} \textbf{\bibinfo{volume}{96}},
  \bibinfo{pages}{012003} (\bibinfo{year}{2006}), \eprint{hep-ph/0511063}.

\bibitem[{\citenamefont{Chetyrkin and Steinhauser}(1997)}]{Chetyrkin:1997vj}
\bibinfo{author}{\bibfnamefont{K.~G.} \bibnamefont{Chetyrkin}}
  \bibnamefont{and}
  \bibinfo{author}{\bibfnamefont{M.}~\bibnamefont{Steinhauser}},
  \bibinfo{journal}{Phys. Lett.} \textbf{\bibinfo{volume}{B408}},
  \bibinfo{pages}{320} (\bibinfo{year}{1997}), \eprint{hep-ph/9706462}.

\bibitem[{\citenamefont{Fleischer and Jegerlehner}(1981)}]{Fleischer:1980ub}
\bibinfo{author}{\bibfnamefont{J.}~\bibnamefont{Fleischer}} \bibnamefont{and}
  \bibinfo{author}{\bibfnamefont{F.}~\bibnamefont{Jegerlehner}},
  \bibinfo{journal}{Phys. Rev.} \textbf{\bibinfo{volume}{D23}},
  \bibinfo{pages}{2001} (\bibinfo{year}{1981}).

\bibitem[{\citenamefont{Bardin et~al.}(1991)\citenamefont{Bardin, Vilensky, and
  Khristova}}]{Bardin:1990zj}
\bibinfo{author}{\bibfnamefont{D.~{\relax Yu}.} \bibnamefont{Bardin}},
  \bibinfo{author}{\bibfnamefont{B.~M.} \bibnamefont{Vilensky}},
  \bibnamefont{and} \bibinfo{author}{\bibfnamefont{P.~K.}
  \bibnamefont{Khristova}}, \bibinfo{journal}{Sov. J. Nucl. Phys.}
  \textbf{\bibinfo{volume}{53}}, \bibinfo{pages}{152} (\bibinfo{year}{1991}),
  \bibinfo{note}{[Yad. Fiz.53,240(1991)]}.

\bibitem[{\citenamefont{Dabelstein and Hollik}(1992)}]{Dabelstein:1991ky}
\bibinfo{author}{\bibfnamefont{A.}~\bibnamefont{Dabelstein}} \bibnamefont{and}
  \bibinfo{author}{\bibfnamefont{W.}~\bibnamefont{Hollik}},
  \bibinfo{journal}{Z. Phys.} \textbf{\bibinfo{volume}{C53}},
  \bibinfo{pages}{507} (\bibinfo{year}{1992}).

\bibitem[{\citenamefont{Kniehl}(1992)}]{Kniehl:1991ze}
\bibinfo{author}{\bibfnamefont{B.~A.} \bibnamefont{Kniehl}},
  \bibinfo{journal}{Nucl. Phys.} \textbf{\bibinfo{volume}{B376}},
  \bibinfo{pages}{3} (\bibinfo{year}{1992}).

\bibitem[{\citenamefont{Mihaila et~al.}(2015)\citenamefont{Mihaila, Schmidt,
  and Steinhauser}}]{Mihaila:2015lwa}
\bibinfo{author}{\bibfnamefont{L.}~\bibnamefont{Mihaila}},
  \bibinfo{author}{\bibfnamefont{B.}~\bibnamefont{Schmidt}}, \bibnamefont{and}
  \bibinfo{author}{\bibfnamefont{M.}~\bibnamefont{Steinhauser}},
  \bibinfo{journal}{Phys. Lett.} \textbf{\bibinfo{volume}{B751}},
  \bibinfo{pages}{442} (\bibinfo{year}{2015}), \eprint{1509.02294}.

\end{thebibliography}

\end{document}